\documentclass[sigconf,authorversion]{acmart} %
\AtBeginDocument{%
  }

\copyrightyear{2025}
\acmYear{2025}
\setcopyright{cc}
\setcctype{by}
\acmConference[VRST '25]{31st ACM Symposium on Virtual Reality Software and Technology}{November 12--14, 2025}{Montreal, QC, Canada}
\acmBooktitle{31st ACM Symposium on Virtual Reality Software and Technology (VRST '25), November 12--14, 2025, Montreal, QC, Canada}
\acmDOI{10.1145/3756884.3766033}
\acmISBN{979-8-4007-2118-2/2025/11}

\graphicspath{{figs/}{./}}

\usepackage[nolist,nohyperlinks]{acronym}
\usepackage[capitalize]{cleveref}

\begin{document}

\title{IGUANA: Immersive Guidance, Navigation, and Control for Consumer UAV}

\author{Victor Victor}
\authornote{Both authors contributed equally to this research.}
\orcid{0009-0005-2434-6318}
\affiliation{%
  \department{Chair of Software Technology}
  \institution{Technische Universität Dresden}
  \city{Dresden}
  \country{Germany}
}
\email{victor.victor@tu-dresden.de}

\author{Tania Krisanty}
\authornotemark[1]
\orcid{0009-0000-2369-7371}
\affiliation{%
  \department{Chair of Computer Graphics and Visualization}
  \institution{Technische Universität Dresden}
  \city{Dresden}
  \country{Germany}
}
\email{tania.krisanty@tu-dresden.de}

\author{Matthew McGinity}
\orcid{0000-0002-8923-6284}
\affiliation{%
  \department{Immersive Experience Lab}
  \institution{Technische Universität Dresden}
  \city{Dresden}
  \country{Germany}
}
\email{matthew.mcginity@tu-dresden.de}

\author{Stefan Gumhold}
\orcid{0000-0003-2467-5734}
\affiliation{%
  \department{Chair of Computer Graphics and Visualization}
  \institution{Technische Universität Dresden}
  \city{Dresden}
  \country{Germany}
}
\email{stefan.gumhold@tu-dresden.de}

\author{Uwe Aßmann}
\orcid{0000-0002-3513-6448}
\affiliation{%
  \department{Chair of Software Technology}
  \institution{Technische Universität Dresden}
  \city{Dresden}
  \country{Germany}
}
\email{uwe.assmann@tu-dresden.de}

\begin{abstract}
As the markets for unmanned aerial vehicles (UAVs) and mixed reality (MR) headsets continue to grow, recent research has increasingly explored their integration, which enables more intuitive, immersive, and situationally aware control systems.
We present IGUANA, an MR-based immersive guidance, navigation, and control system for consumer UAVs.
IGUANA introduces three key elements beyond conventional control interfaces: (1)~a 3D terrain map interface with draggable waypoint markers and live camera preview for high-level control, (2)~a novel spatial control metaphor that uses a virtual ball as a physical analogy for low-level control, and (3)~a spatial overlay that helps track the UAV when it is not visible with the naked eye or visual line of sight is interrupted.
We conducted a user study to evaluate our design, both quantitatively and qualitatively, and found that (1)~the 3D map interface is intuitive and easy to use, relieving users from manual control and suggesting improved accuracy and consistency with lower perceived workload relative to conventional dual-stick controller, (2)~the virtual ball interface is intuitive but limited by the lack of physical feedback, and (3)~the spatial overlay is very useful in enhancing the users' situational awareness.
\end{abstract}



\begin{teaserfigure}
  \includegraphics[width=\textwidth]{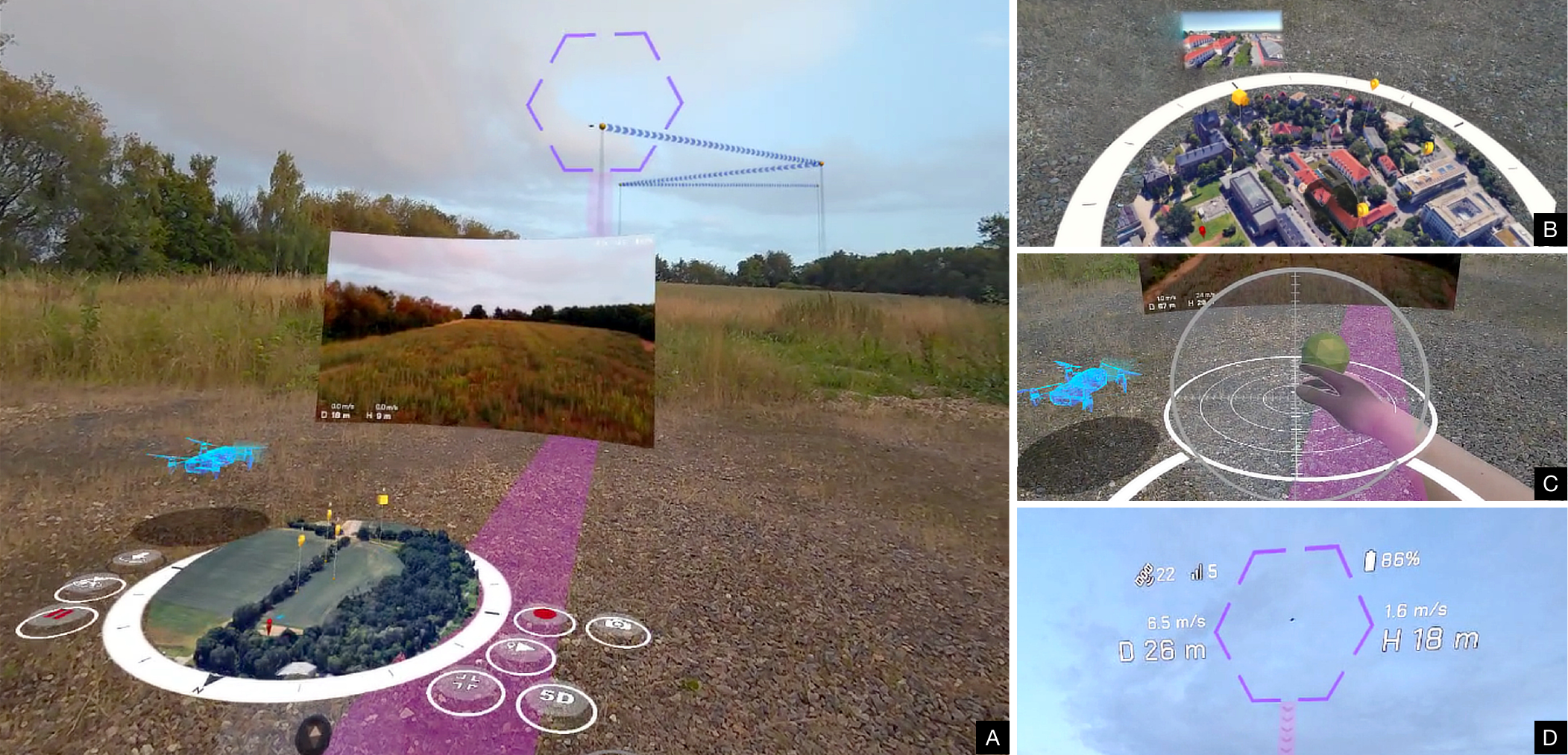}
  \caption{Screenshots of IGUANA captured on Meta Quest 3 headset. (A) Dashboard in 3D map interface mode. (B) High-level control using the 3D map interface. (C) Low-level control using the virtual ball interface. (D) Spatial overlay around the UAV.}
  \Description{Four screenshots of IGUANA running on Meta Quest 3.
  (A) Dashboard in 3D map mode.
  (B) High-level control performed in the 3D map interface.
  (C) Low-level control using the virtual ball interface.
  (D) A spatial overlay around the UAV showing its telemetry.}
  \label{fig:teaser}
\end{teaserfigure}

\maketitle

\begin{acronym}
  \acro{AI}{Artificial Intelligence}
  \acro{AR}{Augmented Reality}
  \acro{CAVE}{Cave Automatic Virtual Environment}
  \acro{ECEF}{Earth-Centered, Earth-Fixed}
  \acro{FPV}{First-Person View}
  \acro{GCS}{Ground Control System}
  \acro{GPS}{Global Positioning System}
  \acro{HMD}{Head-Mounted Display}
  \acro{MR}{Mixed Reality}
  \acro{NASA-TLX}{NASA Task Load Index}
  \acro{NUI}{Natural User Interface}
  \acro{RTMP}{Real-Time Messaging Protocol}
  \acro{SEQ}{Single Ease Question}
  \acro{SUS}{System Usability Scale}
  \acro{TAM}{Technology Acceptance Model}
  \acro{TCP}{Transmission Control Protocol}
  \acro{UAV}{Unmanned Aerial Vehicle}
  \acro{USB}{Universal Serial Bus}
  \acro{VR}{Virtual Reality}
  \acro{WebRTC}{Web Real-Time Communication}
  \acro{WGS84}{World Geodetic System 1984}
  \acro{WIM}{World in Miniature}
  \acro{XR}{Extended Reality}
  \acro{DoF}{degrees of freedom}
\end{acronym}

\section{Introduction}
\label{sec:introduction}

Consumer \acp{UAV} commonly carry cameras for aerial photography, surveying, and monitoring.
Their ability to fly over obstacles, water, and rough terrain makes them well-suited for observing hard-to-reach or large-scale areas, enabling tasks that would otherwise be dangerous, costly, or infeasible. 

While \ac{UAV} motion is intuitive (objects move up/down, turn, and tilt in 3D), producing that motion with conventional controllers is not. 
Users must map intended actions onto limited rotational or translational axes, constrained by the hardware. 
Dual-stick controllers assign two axes per stick, often requiring simultaneous coordination of both.
Users must also operate a gimbal-mounted camera, creating additional complexity.
Even basic tasks demand managing both vehicle and camera orientation, shifting attention from \textit{what to do} to \textit{how to do} it.

Higher-level autonomy, such as waypoint missions, reduces manual navigation by following user-defined \ac{GPS} waypoints.
However, authoring missions remains tedious and error-prone since users still have to manually enter coordinates and altitudes for each waypoint.
As fine-tuning at each waypoint is still common, configuring camera orientation (e.g., pitch) and working without immediate contextual feedback or previews often forces reliance on external tools like Google Earth~\cite{googleGoogleEarth}.

Maintaining a visual line of sight further conflicts with monitoring a controller-mounted live camera feed, increasing cognitive load and reducing situational awareness~\cite{atanasyanAugmentedRealityBasedDrone2023}.
Novices are especially affected by this problem.
While \ac{FPV} goggles offer immersive video, they also limit situational awareness~\cite{sautenkovFlightARARFlight2024}.

These challenges present opportunities for \ac{MR} solutions.
Waypoint planning can benefit from \ac{WIM} interface~\cite{stoakleyVirtualRealityWIM1995}, a scaled 3D representation shown to be safer and more efficient than joysticks or 2D touchscreens~\cite{patersonImprovingUsabilityEfficiency2019}.
Prior work often uses \ac{VR} controllers for placement, which can feel cumbersome compared to hand tracking.
Hand-tracking improves waypoint placement~\cite{salunkheIntuitiveHumanDroneCollaborative2025}, contextual previewing, and route planning while reducing reliance on numeric coordinates and external tools.
Headset-based \ac{MR} also yields better virtual object placement experiences than handheld \ac{AR}~\cite{wachterIndoorDronePath2024}.
A 3D user interface can replace dual-stick limitations by unifying two sticks and a dial into a natural, intuitive virtual ball control.
\ac{MR} can also enhance situational awareness with a persistent, headset-visible indicator of the \ac{UAV}'s real-world location, even when out of sight.

In this work, we propose a design for an \ac{MR}-based guidance, navigation, and control system for consumer \acp{UAV}.
We present IGUANA as a proof-of-concept implementation of this design and evaluate its usability and acceptance in a user study.
This study addresses how IGUANA supports task performance in \ac{UAV} control, how users experience IGUANA in terms of usability and satisfaction, and whether IGUANA enhances situational awareness during \ac{UAV} operation.
The IGUANA source code is publicly available to facilitate future research and development \cite{victorIGUANA2025}.

\section{Related Work}
\label{sec:related-work}

\begin{figure*}[t]
  \centering
  \includegraphics[width=\textwidth]{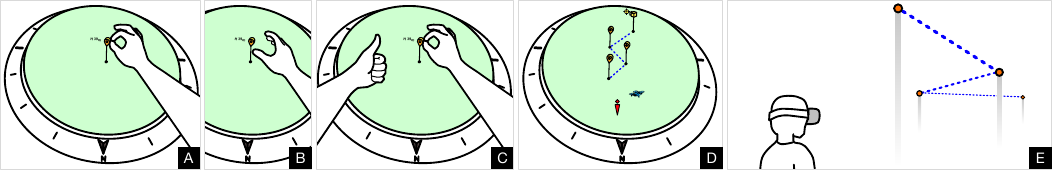}
  \caption{3D map interface. (A) Pinch to select a marker. (B) Release to place the marker. (C) Alternatively, confirm placement with a thumbs-up gesture. (D) Overview with user marker (red), \ac{UAV} (blue), waypoint and virtual camera markers (yellow), and the planned trajectory (dashed line). (E) Spatial overlay of waypoints and the planned trajectory.}
  \Description{Five-panel schematic illustrating waypoint placement in a 3D map interface.
  (A) A hand performs a pinch to pick up a waypoint marker.
  (B) Releasing the pinch drops the marker at the hovered position.
  (C) Alternatively, a thumbs-up gesture confirms the placement.
  (D) Overview: the user's position is shown as a red marker; the UAV is blue; placed waypoints are yellow; a dashed polyline previews the planned trajectory connecting waypoints.
  (E) The same waypoints and dashed trajectory are rendered as a spatial overlay anchored in the environment.}
  \label{fig:map}
\end{figure*}

\begin{figure}[t]
  \centering
  \includegraphics[width=\columnwidth]{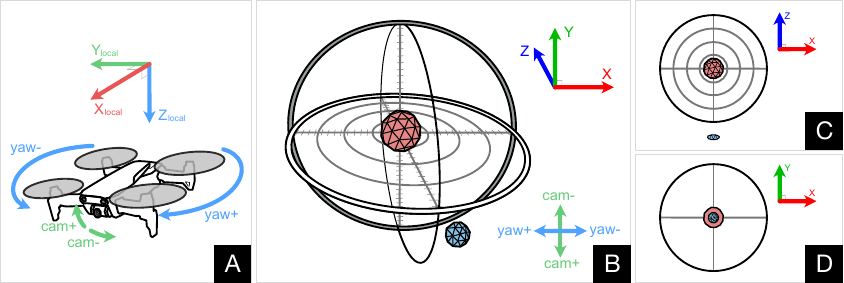}
  \caption{(A) UAV local coordinate system. (B) Virtual ball interface aligned to the world coordinate system; red sphere controls UAV flight direction and speed; blue sphere controls UAV yaw and camera pitch. (C) Top view and (D) front view of the virtual ball interface.}
  \Description{Four panels.
  (A) UAV body-frame axes with X red, Y green, Z blue.
  (B) World-aligned virtual-ball interface where the red sphere sets translational flight direction and the blue sphere adjusts yaw and camera pitch.
  (C) Top view of the interface.
  (D) Front view of the interface.}
  \label{fig:pentacon}
\end{figure}

\paragraph{High-Level Control}

We define high-level control as supervisory or task-level inputs that specify goals, focusing on the desired outcome rather than the specific control actions required to achieve it.
Path planning is a common high-level \ac{UAV} control method, which can be implemented in a grid-based fashion~\cite{wachterIndoorDronePath2024} or using waypoints~\cite{salunkheIntuitiveHumanDroneCollaborative2025,o.VirtualRealityHumanDrone2023,lawtonImmersiveMapsDrone2022,dayaniImmersiveOperationSemiAutonomous2021,patersonImprovingUsabilityEfficiency2019}.
A 3D map~\cite{salunkheIntuitiveHumanDroneCollaborative2025,lawtonImmersiveMapsDrone2022,lawtonCaseEnhancingUAV2022,dayaniImmersiveOperationSemiAutonomous2021,sedlmajerEffectiveRemoteDrone2019} or 3D environment~\cite{wojtkowskiNewExocentricMetaphor2020,betancourtExocentricControlScheme2023}, acting as \ac{WIM}, is employed to provide spatial context.
Some works reconstructed the 3D map from the \ac{UAV}'s camera~\cite{salunkheIntuitiveHumanDroneCollaborative2025,liuAugmentedRealityInteraction2020,morandoSpatialAssistedHumanDrone2024}, while others used pre-existing 3D maps, such as 3D satellite views of the environment~\cite{dayaniImmersiveOperationSemiAutonomous2021} or Mapbox~\cite{sedlmajerEffectiveRemoteDrone2019,mapboxMapboxMapsNavigation2010}.
Voice commands~\cite{helmertDesignEvaluationAR2023,loNaturalHumanDroneInterface2022}, gaze with hand-gestures~\cite{liuAugmentedRealityInteraction2020,eratDroneAugmentedHumanVision2018}, or a combination of both~\cite{huangFlightCameraAction2019} are other methods of high-level control, providing intent- or goal-based control.

Using voice commands requires some understanding of the environment, such as predefined landmarks~\cite{huangFlightCameraAction2019} or object detection~\cite{helmertDesignEvaluationAR2023}, which may not be practical.
On the other hand, some works demonstrate path planning efficiency gains over traditional 2D touchscreen interface~\cite{patersonImprovingUsabilityEfficiency2019,wachterIndoorDronePath2024,o.VirtualRealityHumanDrone2023,emamiUseImmersiveDigital2025}.

In this work, we propose a 3D map interface with waypoint path planning that utilizes photorealistic 3D tiles from Google Earth~\cite{betancourtExocentricControlScheme2023,sedlmajerEffectiveRemoteDrone2019,sinaniDroneTeleoperationInterfaces2025}, providing a rich real-world spatial context for outdoor environments.

\paragraph{Low-Level Control}

Low-level control enables users to directly manipulate the \ac{UAV}'s movement and orientation, providing more granular control over its flight.
The common methods include the use of spatial controllers~\cite{chenMultiVRDigitalTwin2022,leRemoteVisualLineofSight2021,dayaniImmersiveOperationSemiAutonomous2021,asavasirikulkijHumanWorkloadEvaluation2023,betancourtExocentricControlScheme2023,talebVRBasedImmersiveService2023}, dual-stick controllers~\cite{kocerImmersiveViewInterface2022}, or hand-gestures~\cite{morandoSpatialAssistedHumanDrone2024,konstantoudakisDroneControlAR2022,liuAugmentedRealityInteraction2020,eratDroneAugmentedHumanVision2018,allenspachDesignEvaluationMixed2023}.
Some works have proposed using a car-steering metaphor~\cite{chengDesignNoviceFriendlyDrone2024}, a \ac{VR} Treadmill~\cite{sehadLocomotionBasedUAVControl2023}, body posture and voice~\cite{loNaturalHumanDroneInterface2022}, or hand poses ~\cite{agyemangGestureControlMicrodrone2023}.
Although the car steering or \ac{VR} Treadmill provides an intuitive and arguably precise control, they are impractical, especially for outdoor usage.
Additionally, the hand poses may add a mental load by requiring users to remember the different gestures needed for each command.

The hand-gesture methods that interact directly with the \ac{UAV} may not be suitable for outdoor use due to the limited interaction distance and the need for a line-of-sight~\cite{morandoSpatialAssistedHumanDrone2024,liuAugmentedRealityInteraction2020,eratDroneAugmentedHumanVision2018}.
In contrast, hand-gesture methods that enable the \ac{UAV} to mimic the hand's pose~\cite{konstantoudakisDroneControlAR2022,allenspachDesignEvaluationMixed2023} are better suited for outdoor use.
The disadvantage of hand-gesture control is the lack of physical feedback, which may affect the control precision and user confidence~\cite{sehadLocomotionBasedUAVControl2023,ibrahimovDronePickObjectPicking2019,yashinAeroVrVirtualRealitybased2019}.

In this work, we propose a virtual ball interface that enables more natural and intuitive control, acting like a joystick with a rest position when not in use~\cite{konstantoudakisDroneControlAR2022,allenspachDesignEvaluationMixed2023}.

\paragraph{Spatial Overlay}

\citet{atanasyanAugmentedRealityBasedDrone2023} made a significant contribution to the field of spatial overlays for \ac{UAV} operation, with \ac{UAV} attitude and position indicators.
When waypoint control is involved, previous studies have demonstrated the benefits of displaying the planned waypoint path in a spatial overlay~\cite{emamiUseImmersiveDigital2025,helmertDesignEvaluationAR2023,walkerCommunicatingRobotMotion2018,zollmannFlyARAugmentedReality2014}.

In this work, we propose a spatial overlay to enhance situational awareness, building on previous work that shows the position of the \ac{UAV} in the air~\cite{atanasyanAugmentedRealityBasedDrone2023,zollmannFlyARAugmentedReality2014}, attitude and telemetry information~\cite{atanasyanAugmentedRealityBasedDrone2023,liuAugmentedRealityInteraction2020}, as well as the waypoint path~\cite{emamiUseImmersiveDigital2025,helmertDesignEvaluationAR2023,walkerCommunicatingRobotMotion2018,zollmannFlyARAugmentedReality2014}.

\section{IGUANA}
\label{sec:iguana}

\subsection{Design}

IGUANA features three key components: high-level control via 3D map interface, low-level control via virtual ball interface, and enhanced situational awareness via spatial overlay.

\subsubsection{High-Level Control via 3D Map Interface}

High-level control refers to planning and executing waypoint missions for the \ac{UAV}.
This is achieved through a 3D map interface that displays a spatially accurate, scaled-down 3D reconstruction of the real world.
The 3D map is displayed in a floating circular cutout in front of the user, which can be panned, rotated, and zoomed (see~\cref{fig:teaser}-A).
On top of the map, the user can place, move, and orient waypoint markers using pinch gestures (see~\cref{fig:map}-A) and release gestures (see~\cref{fig:map}-B).
An additional thumbs-up gesture (with the other hand) is added to help release the marker more precisely (see~\cref{fig:map}-C).
There is also a virtual camera marker (see~\cref{fig:map}-D, top-most marker), in addition to the regular waypoint markers, which displays a live preview of the scene to assist with pose adjustment (see~\cref{fig:teaser}-B, floating window at the top).
The virtual camera marker has an additional reticle helper that can be pulled to adjust the \ac{UAV}'s yaw and camera pitch.
When a hand is near a waypoint or virtual camera marker, height information is displayed to assist with altitude planning, particularly to avoid collisions with the terrain (see~\cref{fig:map}-A, B, and C).
A \ac{UAV} marker and a user marker are also displayed on the map (see~\cref{fig:map}-D, two bottom-most markers).

\begin{figure}[t]
  \centering
  \includegraphics[width=\columnwidth]{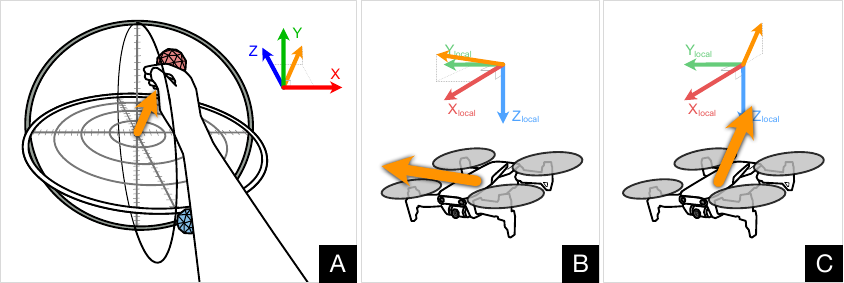}
  \caption{(A) Push on the virtual ball diagonally toward the user's forward upper right. Resulting motion in (B) drone-centric and (C) user-centric control modes.}
  \Description{Three panels.
  (A) A diagonal push on the virtual ball toward the user's forward upper right.
  (B) The resulting motion under drone-centric mode.
  (C) The resulting motion under user-centric mode.}
  \label{fig:drone-vs-user-centric}
\end{figure}

\subsubsection{Low-Level Control via Virtual Ball Interface}

Low-level control refers to direct manipulation of a \ac{UAV}'s attitude (roll, pitch, yaw, and thrust) to achieve a desired flight direction.
To abstract this complexity, we introduce a control metaphor based on a virtual ball, serving as a physical analogy for the \ac{UAV}.
The interface consists of two spheres: a big sphere for translational movement (forward/backward, left/right, up/down) and a small sphere for yaw and camera pitch (see~\cref{fig:pentacon}-A and B, big sphere is colored red, small sphere is colored blue).
The big sphere has 3\ac{DoF} movement, bounded within a larger transparent sphere, while the small sphere has 2\ac{DoF}, constrained to move along a vertical circular path (see~\cref{fig:pentacon}-C and D).
To initiate translational control, the user reaches out and pushes the big sphere in the desired direction.
The small sphere, however, is manipulated using a pinch gesture.
It is possible to control both spheres simultaneously using both hands, combining translational and rotational control.

We implemented two interchangeable control schemes for the virtual ball interface: the \textit{drone-centric} and \textit{user-centric} control.
In the \textit{drone-centric} control, the virtual ball interface is aligned with the \ac{UAV}'s local coordinate system.
Pushing the big sphere in a certain direction will cause the \ac{UAV} to move in the same direction (see~\cref{fig:drone-vs-user-centric}-A and B), as it places the user in a virtual position behind the \ac{UAV}.
The \textit{user-centric} control, on the other hand, aligns the virtual ball interface with the world coordinate system.
Pushing the big sphere in a certain direction will cause the \ac{UAV} to move in that direction relative to the user, regardless of the \ac{UAV}'s current orientation (see~\cref{fig:drone-vs-user-centric}-A and C).
This control scheme is implemented to help reduce the cognitive load associated with heading-relative controls, especially when controlling the \ac{UAV} when it is positioned in front of the user, rather than through the camera stream.

\begin{figure}[t]
  \centering
  \includegraphics[width=\columnwidth]{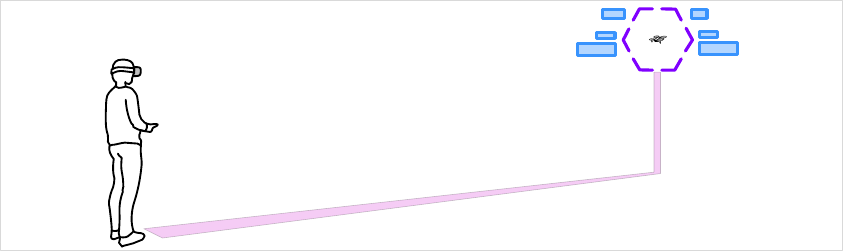}
  \caption{Spatial overlay with horizontal and vertical leading lines (magenta), UAV frame (purple), and telemetry (blue).}
  \Description{Sketch showing a third-person view of a user observing the spatial overlay.}
  \label{fig:spatial-overlay}
\end{figure}

\subsubsection{Enhanced Situational Awareness via Spatial Overlay}
\label{ssec:uav-position-overlay}

Spatial overlay refers to augmenting the real world with virtual information anchored to the \ac{UAV}'s actual position.
This helps users maintain awareness of the \ac{UAV}'s position, especially when it is too far away or beyond the line of sight (see~\cref{fig:spatial-overlay}, purple frame).
To help track the \ac{UAV}'s position, we added horizontal leading lines that extend from the user's feet to the \ac{UAV}'s ground position and vertical leading lines that extend from the ground to the \ac{UAV}'s altitude (see~\cref{fig:spatial-overlay}, magenta lines).
The overlay displays its telemetry information, such as ground distance, ground speed, altitude, vertical speed, \ac{GPS} signal strength, and battery level (see~\cref{fig:teaser}-D and \ref{fig:spatial-overlay}, blue panels).
As the user plans the waypoint missions with the 3D map interface (see~\cref{fig:map}-D), we also visualize the planned waypoint path in the real world, providing context for the \ac{UAV}'s intended trajectory (see~\cref{fig:map}-E).
\citet{walkerCommunicatingRobotMotion2018} found that this visualization effectively communicates the intended flight path.

\subsection{Implementation}
\label{subsec:implementation}

\begin{figure}[t]
  \centering
  \includegraphics[width=\columnwidth]{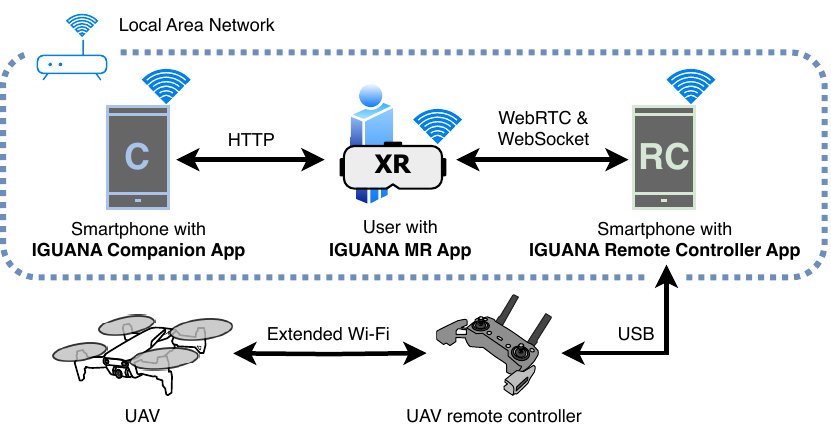}
  \caption{IGUANA system architecture.}
  \Description{Diagram showing IGUANA system architecture.}
  \label{fig:system-architecture}
\end{figure}

Based on the system architecture depicted in~\Cref{fig:system-architecture}, we implemented IGUANA using:
Da-Jiang Innovations (DJI) Mavic Air,
Meta Quest 3,
Samsung Galaxy Note20 Ultra for remote controller application,
and iPhone 16 Pro Max for companion application.
IGUANA, however, is designed to be compatible with any consumer \ac{UAV} that supports waypoint missions and telemetry via a mobile SDK.

\subsubsection{Mixed Reality Application}

The \ac{MR} application is built using Unity 6 with Meta \ac{XR} SDK v77.
\Cref{fig:teaser}-A shows the dashboard with the 3D map interface and the \ac{UAV} marked with a spatial overlay in the background.

\paragraph{3D Map Interface}

We implemented the 3D map interface using Cesium-based Google's Photorealistic 3D Tiles~\cite{cesiumGoogleMaps} (see~\cref{fig:teaser}-B).
The \ac{UAV} marker updates in real-time, based on its actual \ac{GPS} position and altitude value.
The user marker position is determined by the iPhone's location.
Users can interact with the waypoint and virtual camera markers using pinch gestures for both positioning and orientation adjustments.
The virtual camera marker renders a live view of the 3D tiles environment into the viewport and represents the camera mounted on the \ac{UAV}, with its rotation and field of view constrained to match it.

After placing the waypoints, the user can start the mission by pressing the start mission button.
Waypoint data is then transmitted to the remote controller application, which translates each marker's position and orientation into flight parameters, and then forwards it to the \ac{UAV}.
Each position is converted into \ac{GPS} coordinates and an altitude.
The rotation around the X-axis sets the camera's pitch, while rotation around the Y-axis determines the \ac{UAV}'s heading.
One important note is that the altitude value sent to and received from the \ac{UAV} is defined relative to the takeoff point, not in \ac{WGS84}.
Therefore, it is necessary to first obtain the altitude of the takeoff location by querying the height of the 3D tiles where the \ac{UAV} is located.

\paragraph{Virtual Ball Interface}

The virtual ball interface (see~\cref{fig:teaser}-C) is implemented using Meta \ac{XR} Interaction SDK v77.
To support intuitive interaction, we provide visual and auditory feedback.
When the user pushes the ball, it changes color to indicate that it is being pushed, and reverts to its original color upon release.
Additionally, sound cues vary based on how far the ball is pushed, compensating for the lack of haptic feedback.
The distance of the virtual ball from its resting position is mapped to the \ac{UAV}'s pitch, roll, and thrust.

\paragraph{Spatial Overlay}

To overlay the \ac{UAV}'s position accurately, we use Cesium georeference in real-world scale mode, without loading terrain or imagery tiles.
This enables overlays to be positioned based on the \ac{UAV}'s latitude, longitude, and altitude in the \ac{WGS84} system.
To ensure correct spatial alignment, we first set the user's position as the origin of Cesium's georeference.
At initialization, the user holds an iPhone running the Companion Application until the headset detects the AprilTag.
AprilTag is a robust marker-based tracking system for accurate 6\ac{DoF} pose estimation~\cite{krogiusFlexibleLayoutsFiducial2019}.
The headset computes the AprilTag's pose in its coordinate system and simultaneously queries the iPhone's geographic location.

\begin{figure}[t]
  \centering
  \includegraphics[width=\columnwidth]{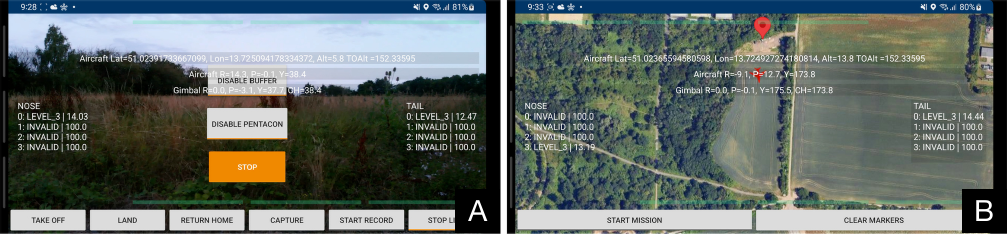}
  \caption{Screenshots from IGUANA remote controller application showing (A) the camera and (B) the map interfaces.}
  \Description{Screenshots from IGUANA remote controller application showing camera interface and map interface.}
  \label{fig:remote-controller-app}
\end{figure}

\begin{figure}[t]
  \centering
  \includegraphics[width=\columnwidth]{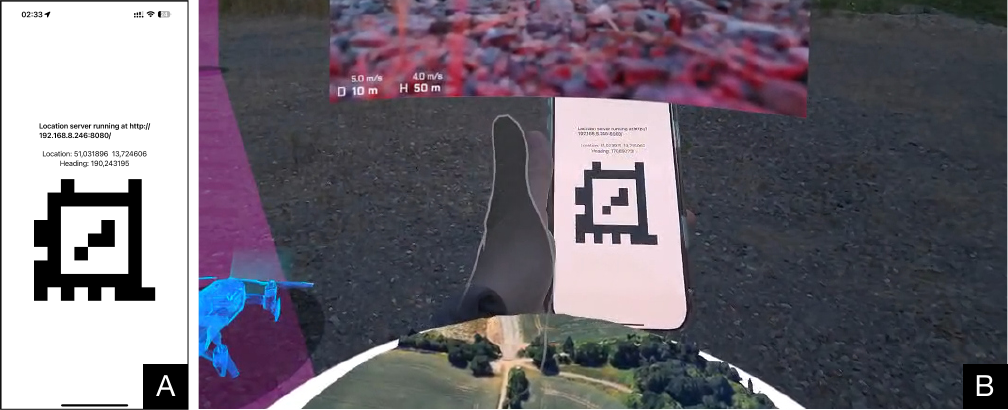}
  \caption{(A) Screenshot of the companion application running on an iPhone. (B) IGUANA MR application scanning the AprilTag displayed on the companion application.}
  \Description{Screenshot of the companion application and screenshot of IGUANA scanning the AprilTag displayed on the companion appplication.}
  \label{fig:companion-app}
\end{figure}

\subsubsection{Remote Controller Application}

The remote controller application is built in Java using the DJI Mobile SDK v4.18.
It connects to the \ac{UAV} remote controller via \ac{USB} and to the headset via Wi-Fi.
It has two main interfaces, switchable by tapping the screen, namely the camera interface (see~\cref{fig:remote-controller-app}-A) and the map interface (see~\cref{fig:remote-controller-app}-B).
The camera interface displays the live feed from the \ac{UAV}'s camera and includes all the important buttons for emergency stop, takeoff, landing, return to home, and a safety override that blocks control input from the virtual ball interface in the \ac{MR} application.
On the other hand, the map interface is useful for showing the \ac{UAV}'s position on the map and monitoring the waypoint mission, as it displays every waypoint marker as it is sent from the headset.

\subsubsection{Companion Application}

The companion application is built in Swift (see~\cref{fig:companion-app}-A and B).
It connects to the headset via Wi-Fi and provides the phone's \ac{GPS} and compass data as IGUANA's position and orientation.
Additionally, it displays an AprilTag, which is used as a pose reference for the headset.

\subsubsection{Communication Protocol}

Communication between the headset and the Android phone is handled via a WebSocket and \ac{WebRTC}, a protocol suggested by previous work~\cite{widiyantiHoloGCSMixedRealitybased2024}.
High-frequency data, such as telemetry updates at 10 Hz, is sent from the phone to the headset via \ac{WebRTC} data channel.
Commands are sent from the headset to the phone via WebSocket, ensuring reliable delivery and correct execution order.
To minimize latency and payload size, data is transmitted as raw bytes, where the first byte indicates the message type, and the number and structure of subsequent bytes vary depending on the type.
This compact format ensures fast and consistent communication for frequent updates.
For real-time preview and situational awareness, the \ac{UAV}'s live camera feed was streamed to the headset via \ac{WebRTC}.
This protocol was preferred over the \ac{UAV}'s built-in \ac{RTMP} streaming due to its lower latency.

\section{User Study}
\label{sec:user-study}

\begin{figure}[t]
  \centering
  \includegraphics[width=\columnwidth]{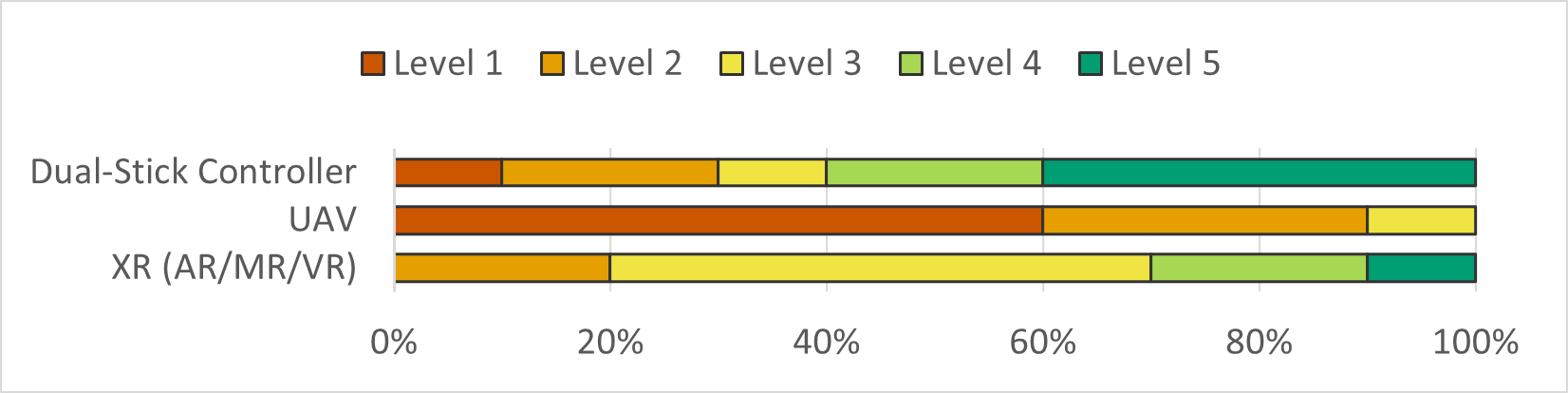}
  \caption{Participants' domain-specific experience level.}
  \Description{Chart showing participants' domain-specific experience level.}
  \label{fig:participants-demography}
\end{figure}

We evaluated IGUANA in a within-subjects study where participants completed the same tasks twice: once with IGUANA and once with a baseline.
For the baseline, referred to as RC, we used the DJI Mavic Air remote controller and DJI GO 4~\cite{djiDownloadCenter} on a Samsung Galaxy S10.
Although IGUANA's 3D map interface is designed for complex environments, such as dense urban terrain, we conducted the study in a safe, open field to minimize risk.

\subsection{Objectives}

The main objectives were to evaluate the usability and effectiveness of IGUANA in comparison to conventional control interfaces for consumer \acp{UAV}. We aimed to assess the following aspects:
(1)~\textbf{Task Performance}: Measure the time taken and accuracy in completing a specific \ac{UAV} control task using both IGUANA and the RC.
(2)~\textbf{User Experience}: Gather qualitative feedback on the overall user experience, including ease of use, intuitiveness, and satisfaction with the control methods.
(3)~\textbf{Situational Awareness}: Evaluate how well IGUANA enhances situational awareness during \ac{UAV} operation.

\begin{figure*}[t]
  \centering
  \begin{minipage}{0.8\textwidth}
    \centering
    \begin{minipage}{\textwidth}
      \centering
      \includegraphics[width=0.095\textwidth]{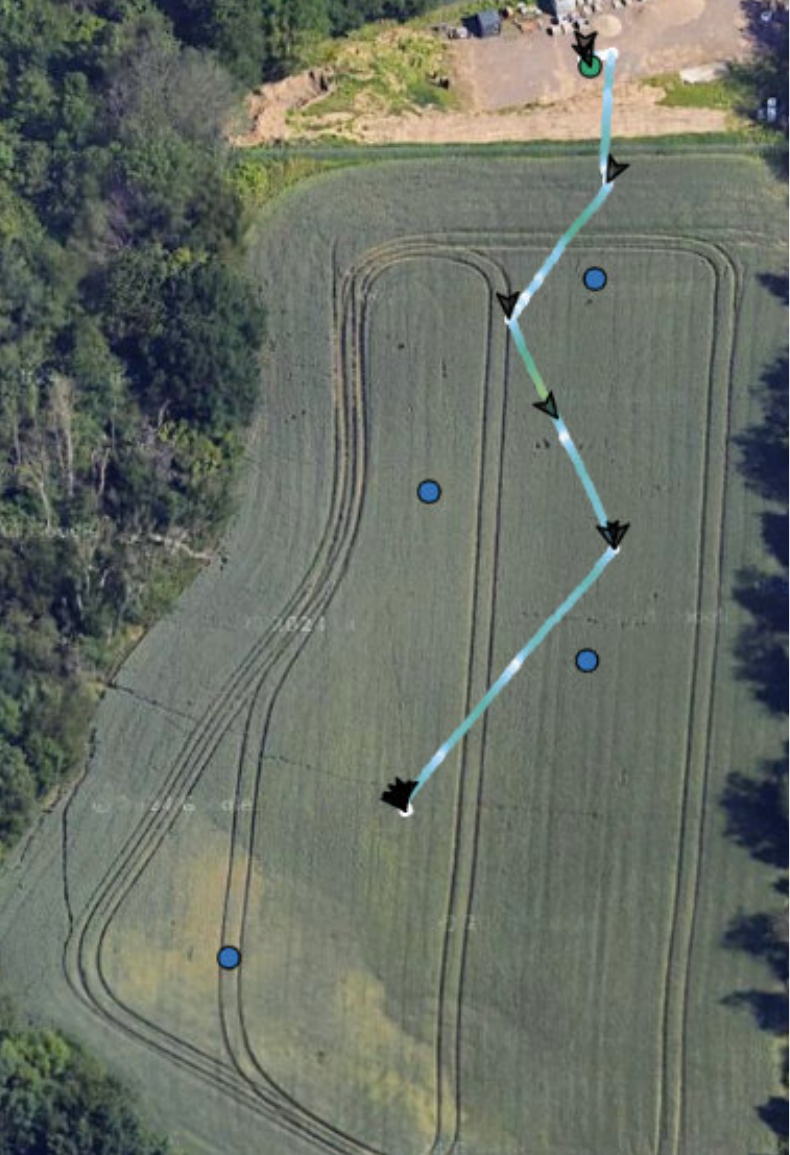}
      \includegraphics[width=0.095\textwidth]{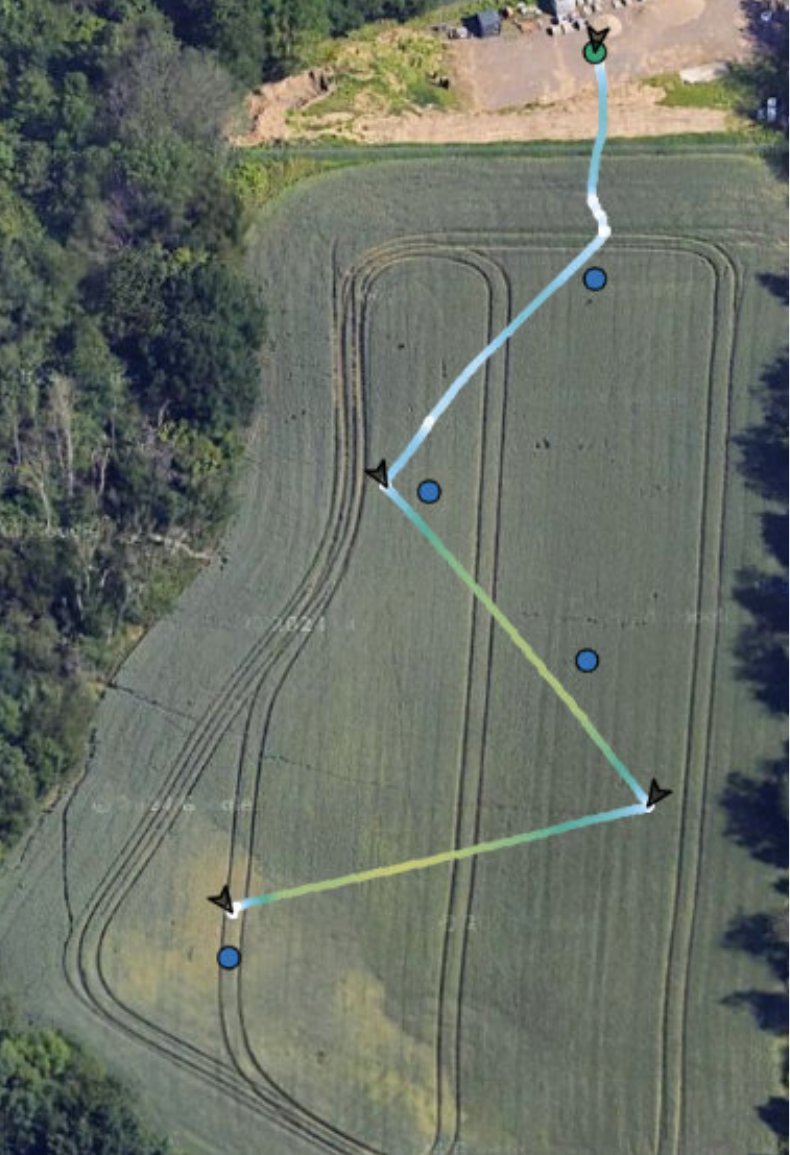}
      \includegraphics[width=0.095\textwidth]{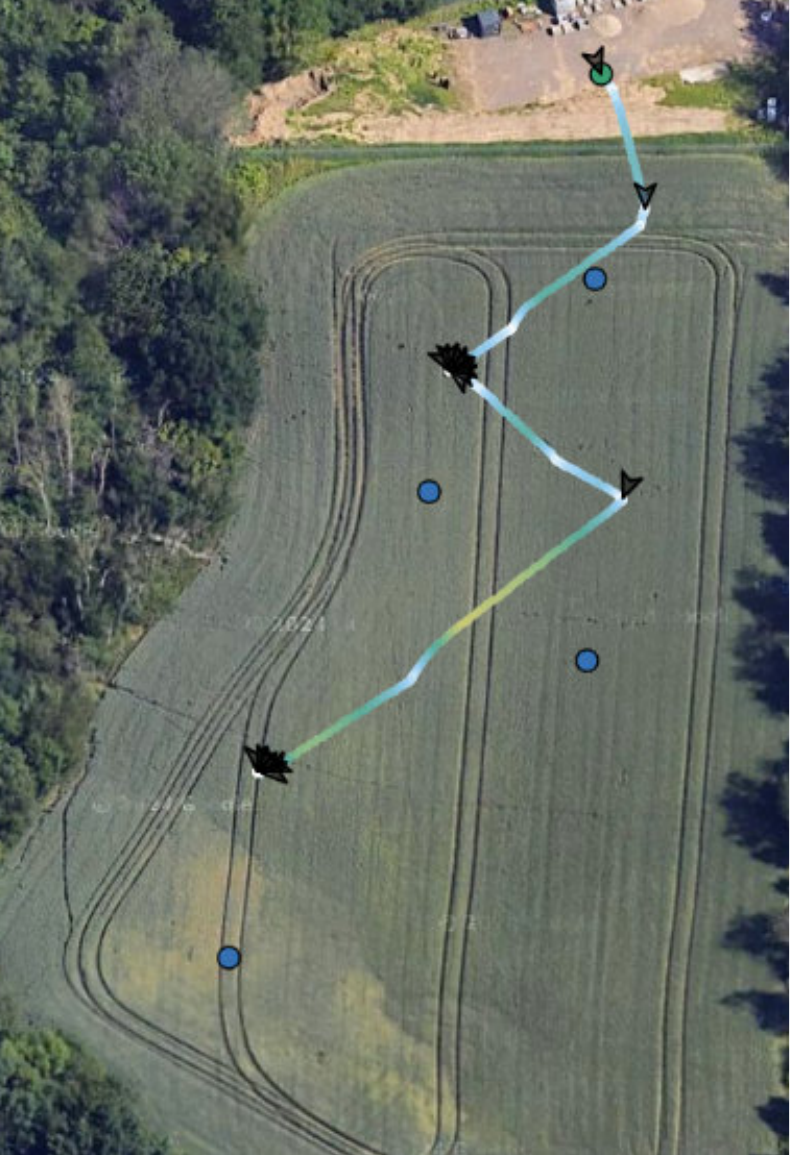}
      \includegraphics[width=0.095\textwidth]{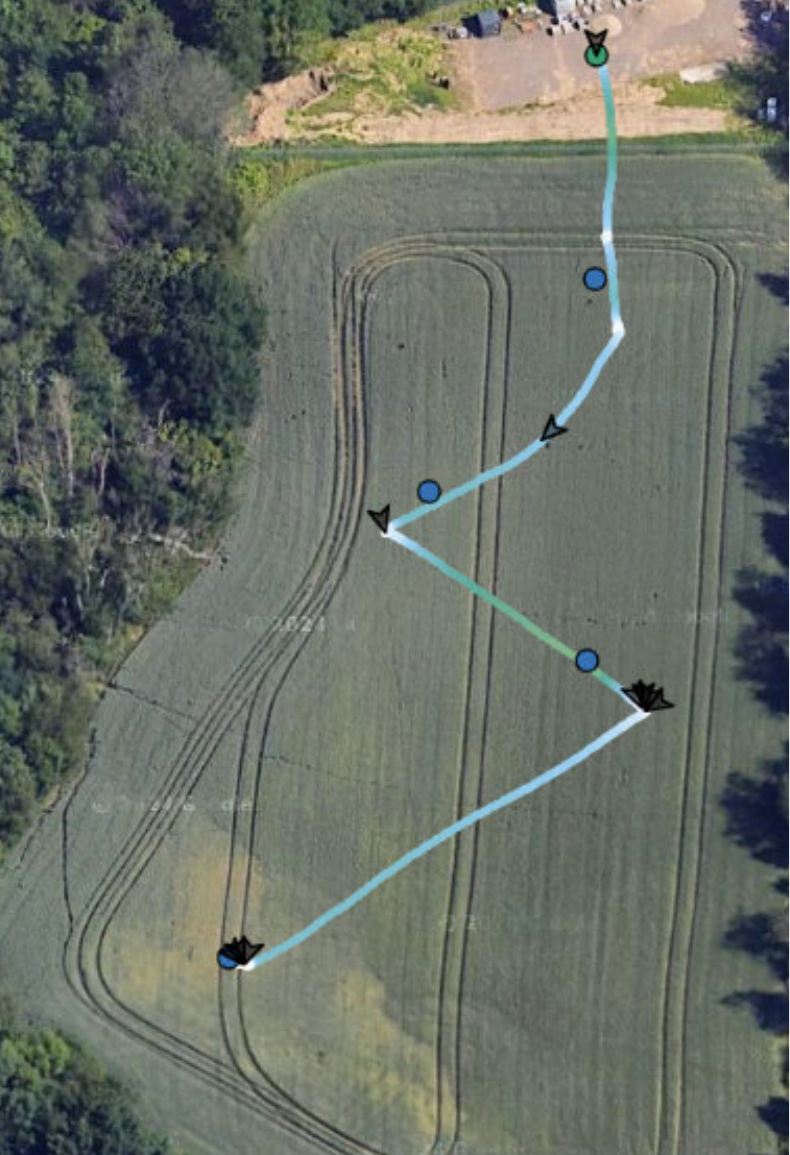}
      \includegraphics[width=0.095\textwidth]{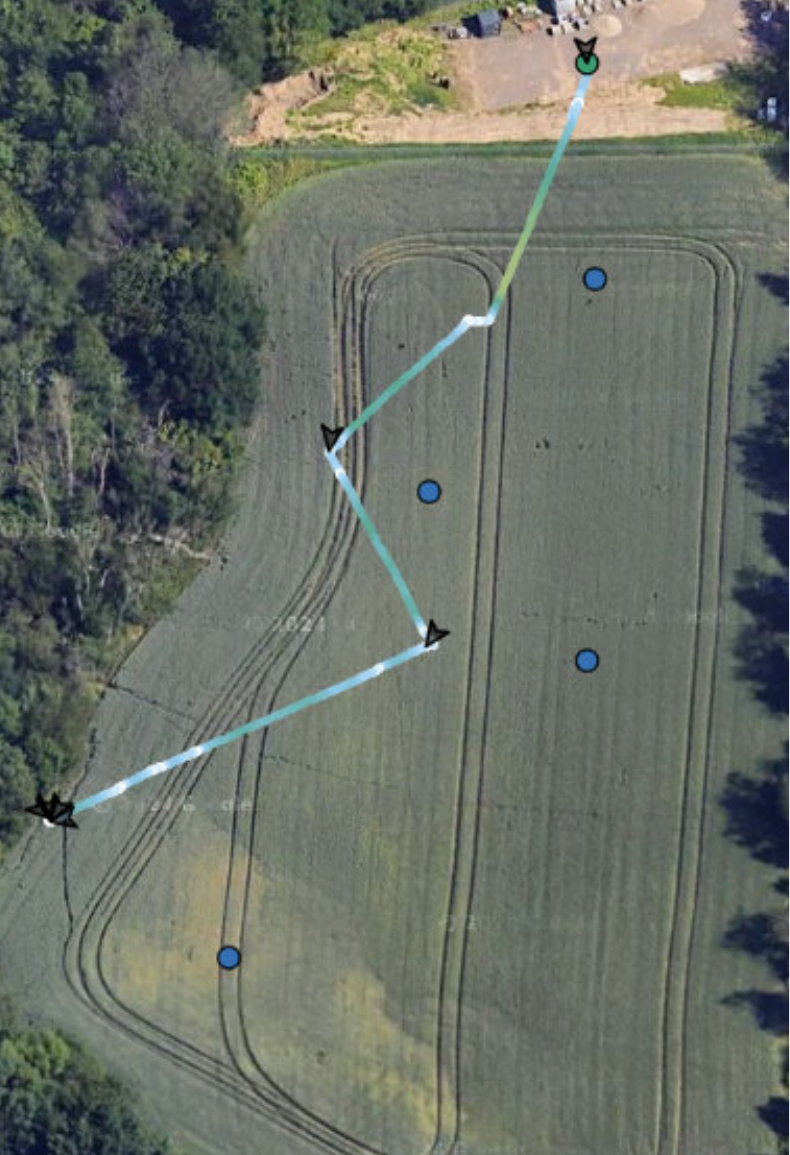}
      \includegraphics[width=0.095\textwidth]{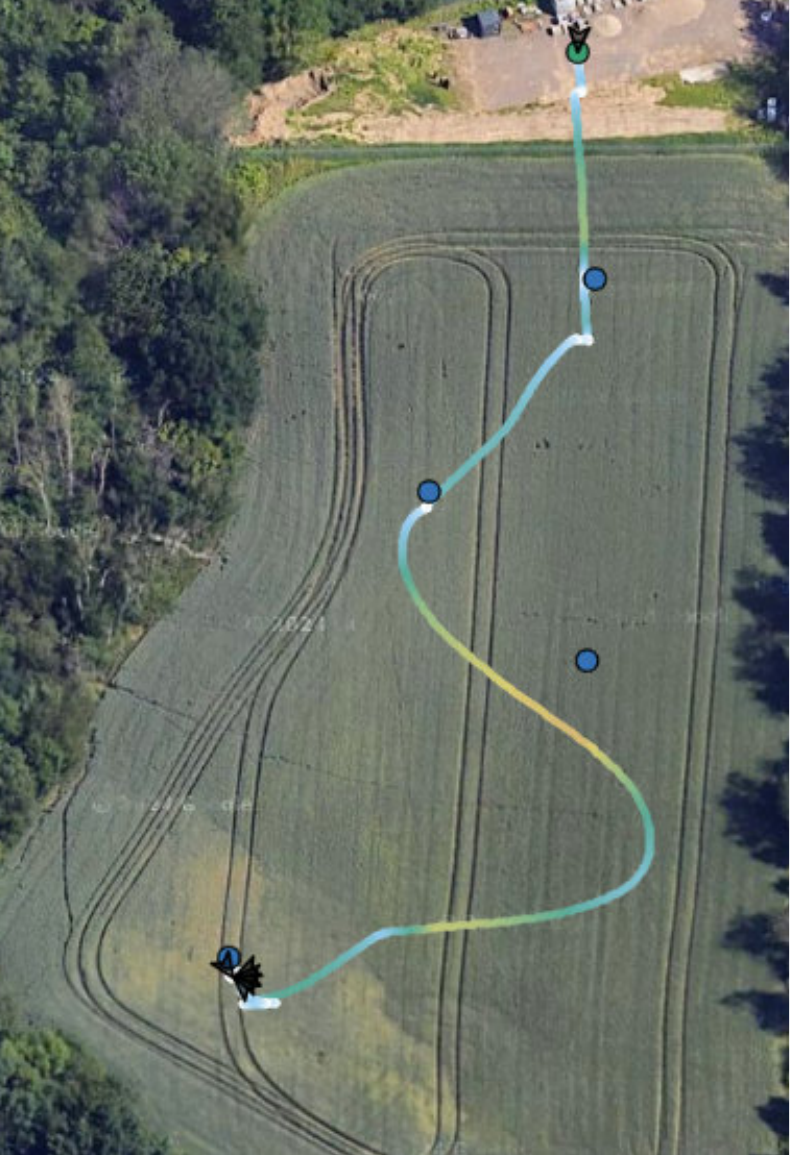}
      \includegraphics[width=0.095\textwidth]{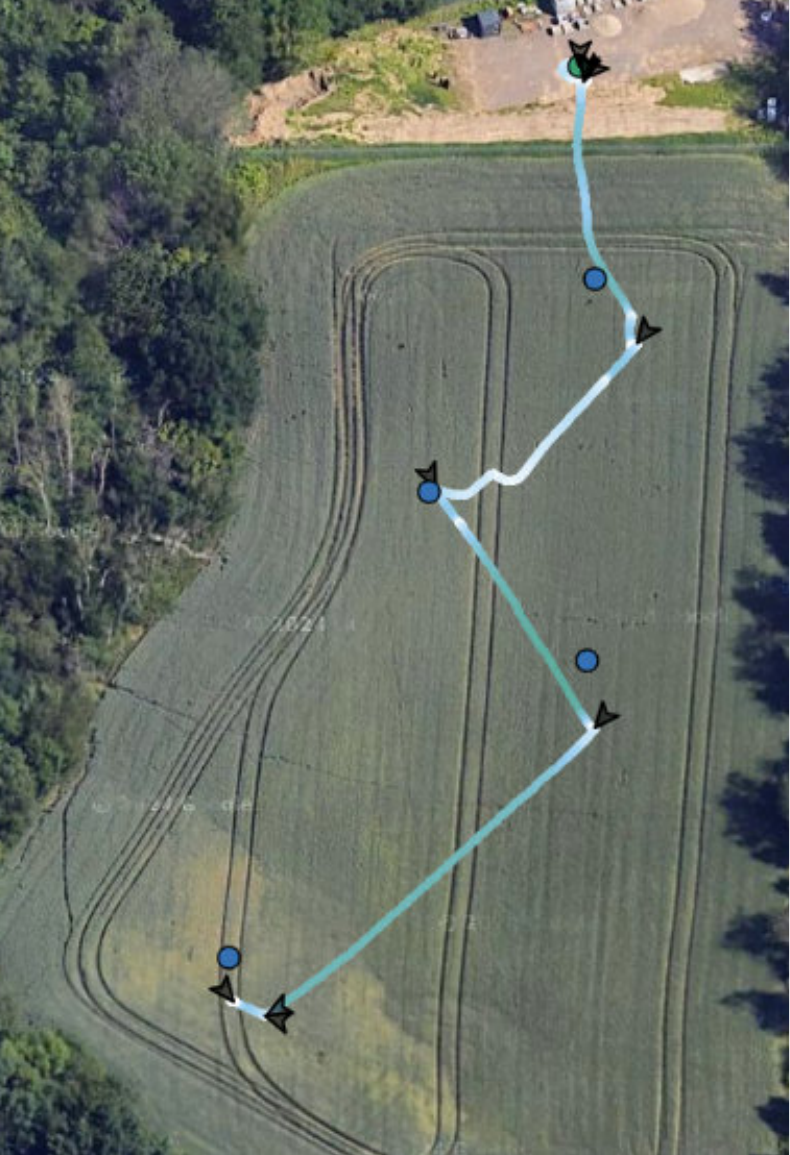}
      \includegraphics[width=0.095\textwidth]{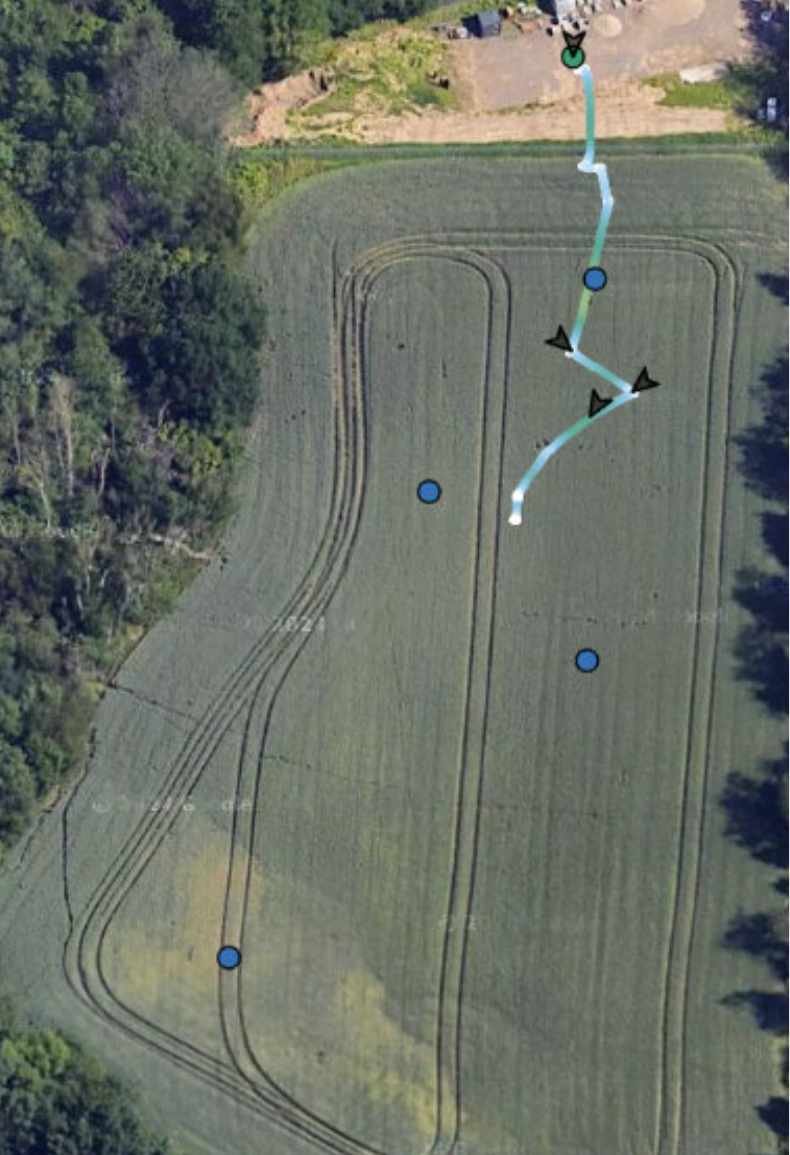}
      \includegraphics[width=0.095\textwidth]{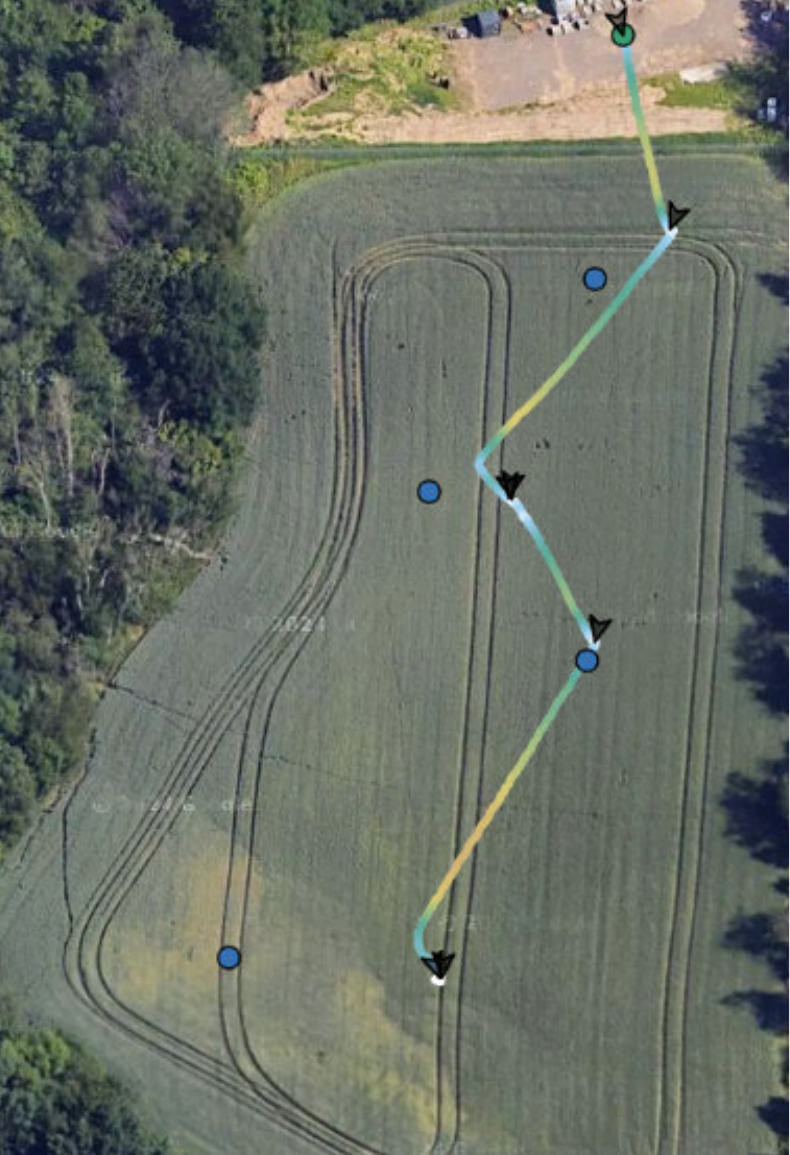}
      \includegraphics[width=0.095\textwidth]{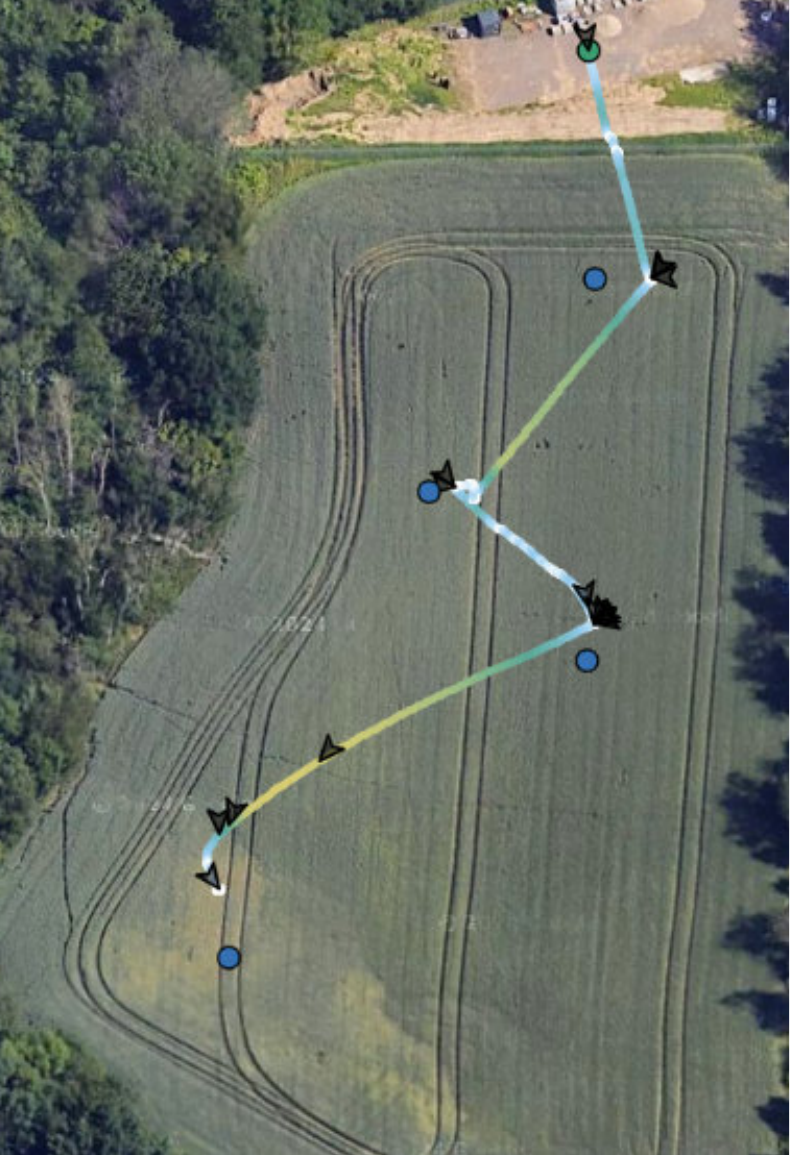}
    \end{minipage}
    \begin{minipage}{\textwidth}
      \centering
      \includegraphics[width=0.095\textwidth]{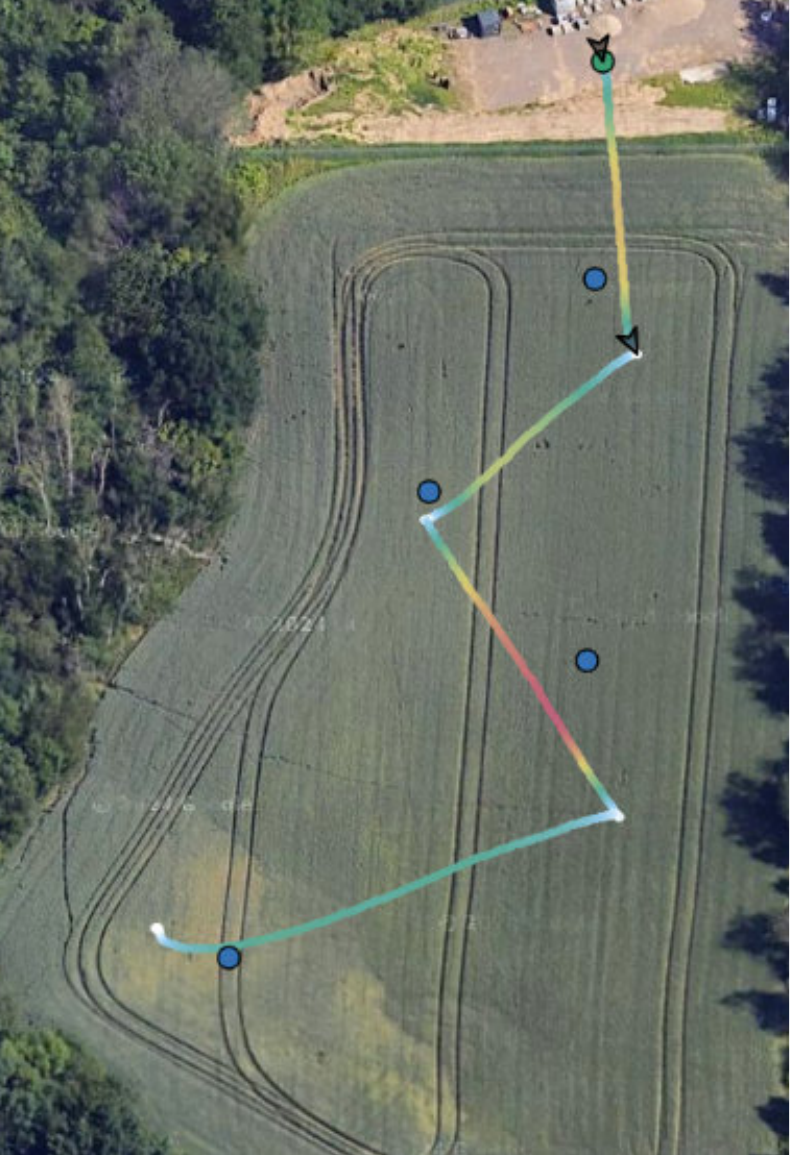}
      \includegraphics[width=0.095\textwidth]{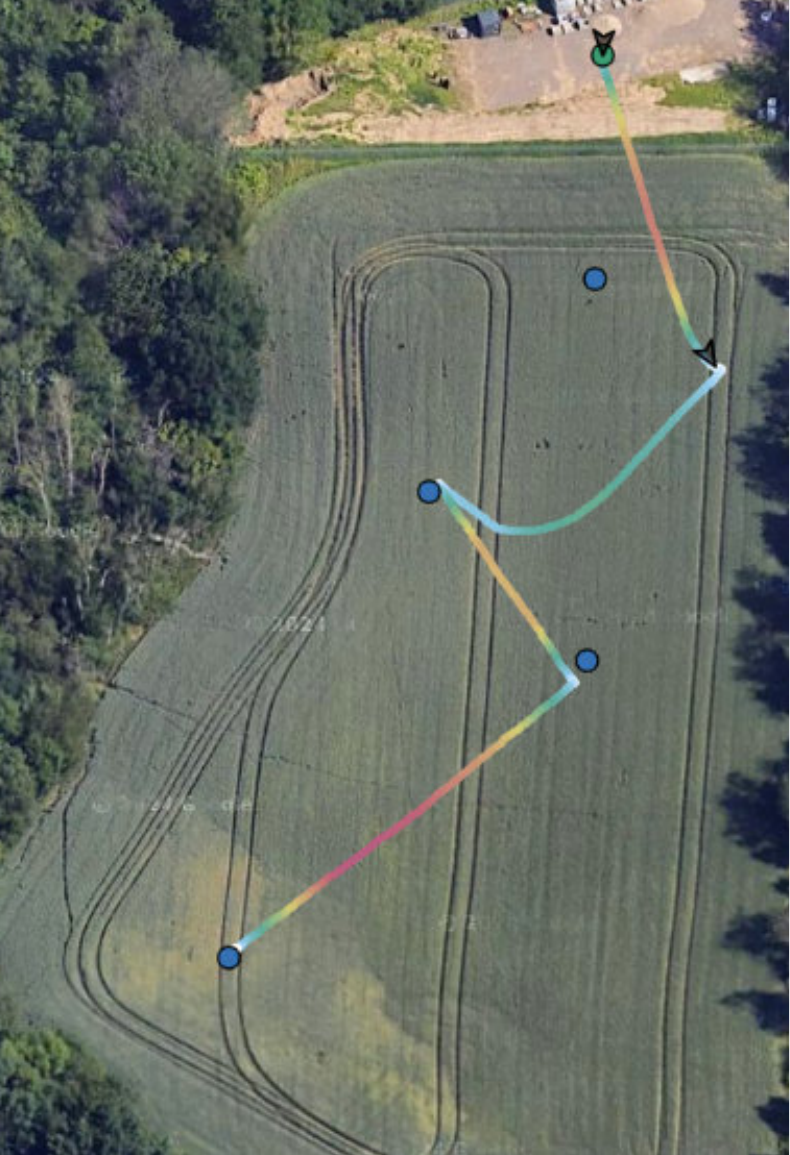}
      \includegraphics[width=0.095\textwidth]{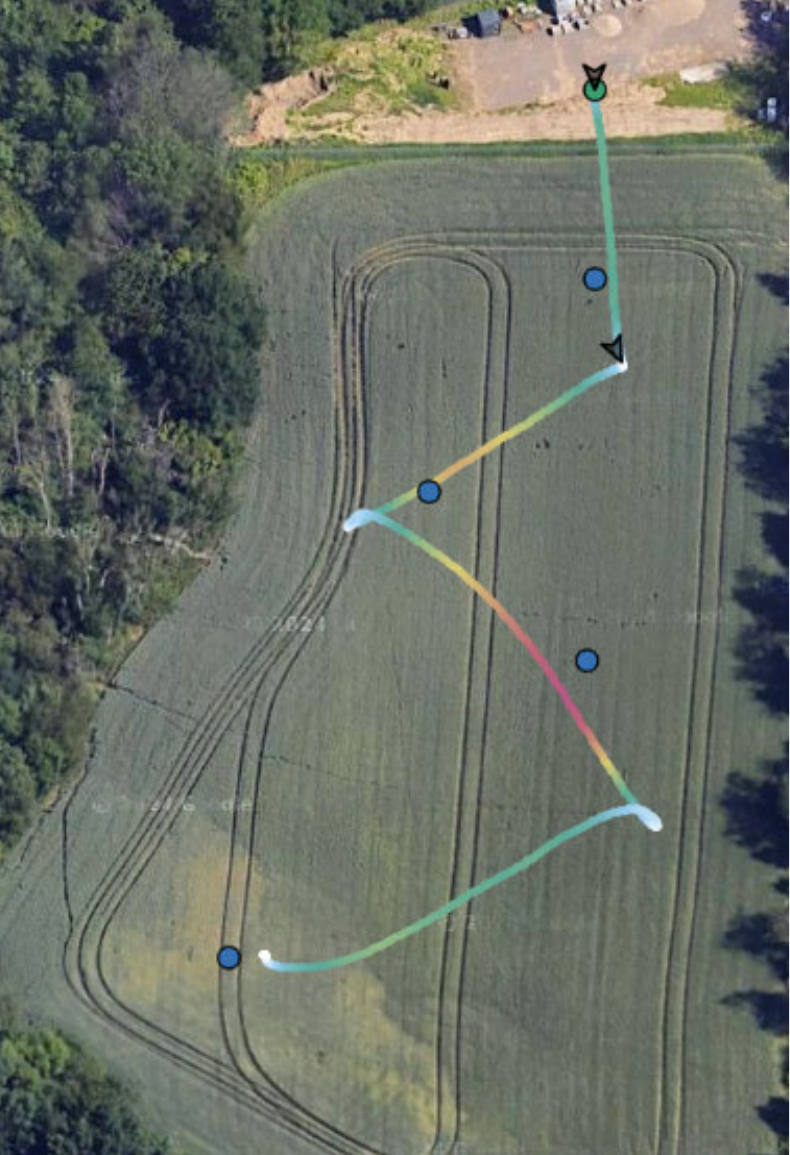}
      \includegraphics[width=0.095\textwidth]{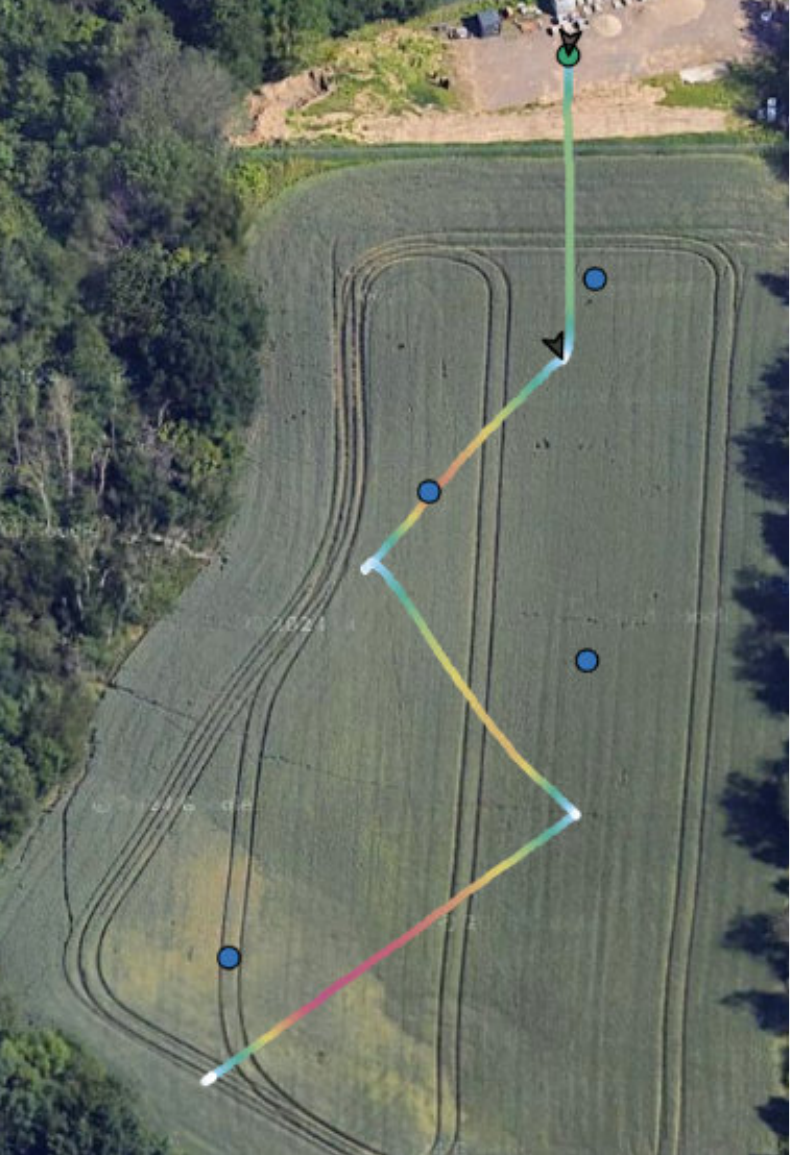}
      \includegraphics[width=0.095\textwidth]{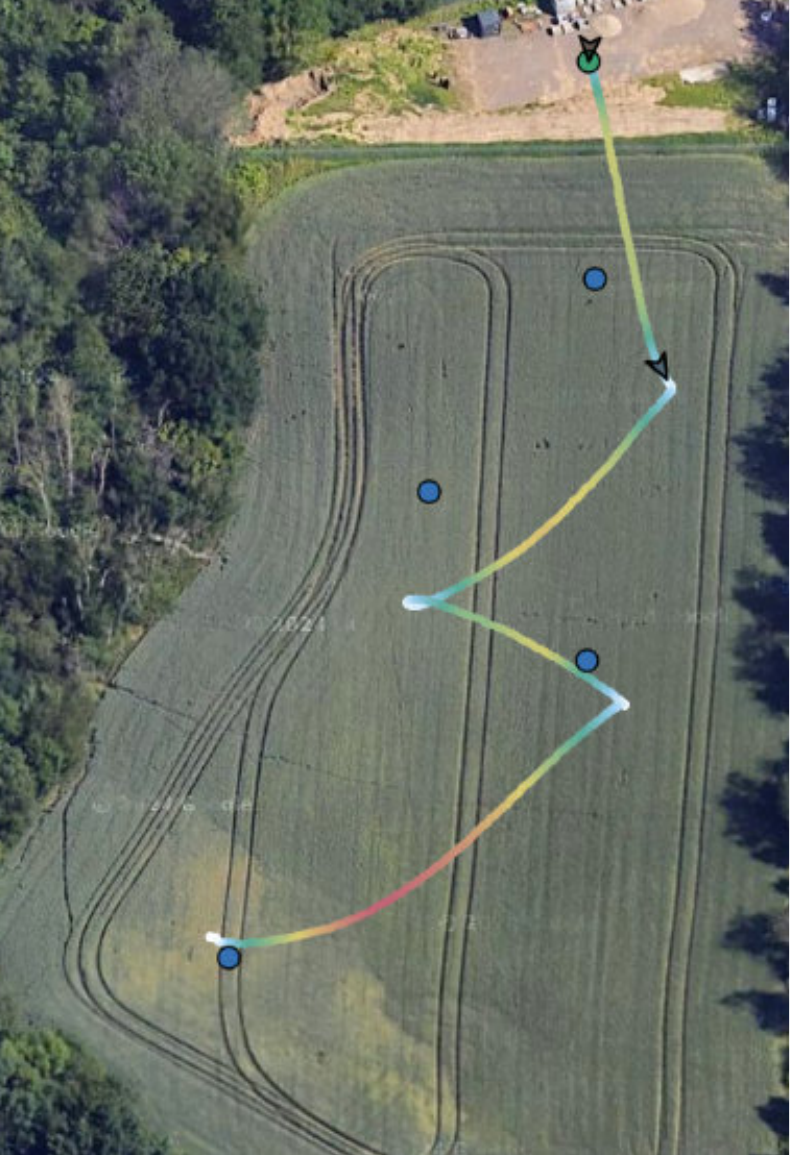}
      \includegraphics[width=0.095\textwidth]{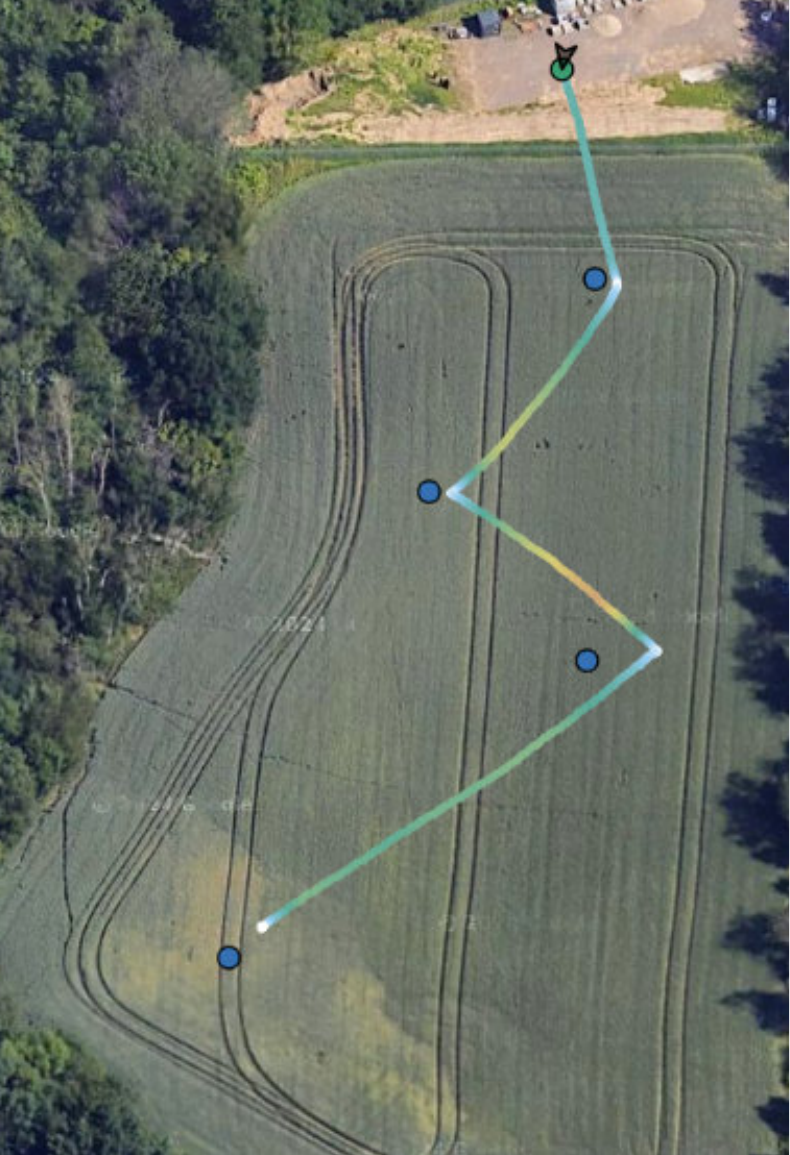}
      \includegraphics[width=0.095\textwidth]{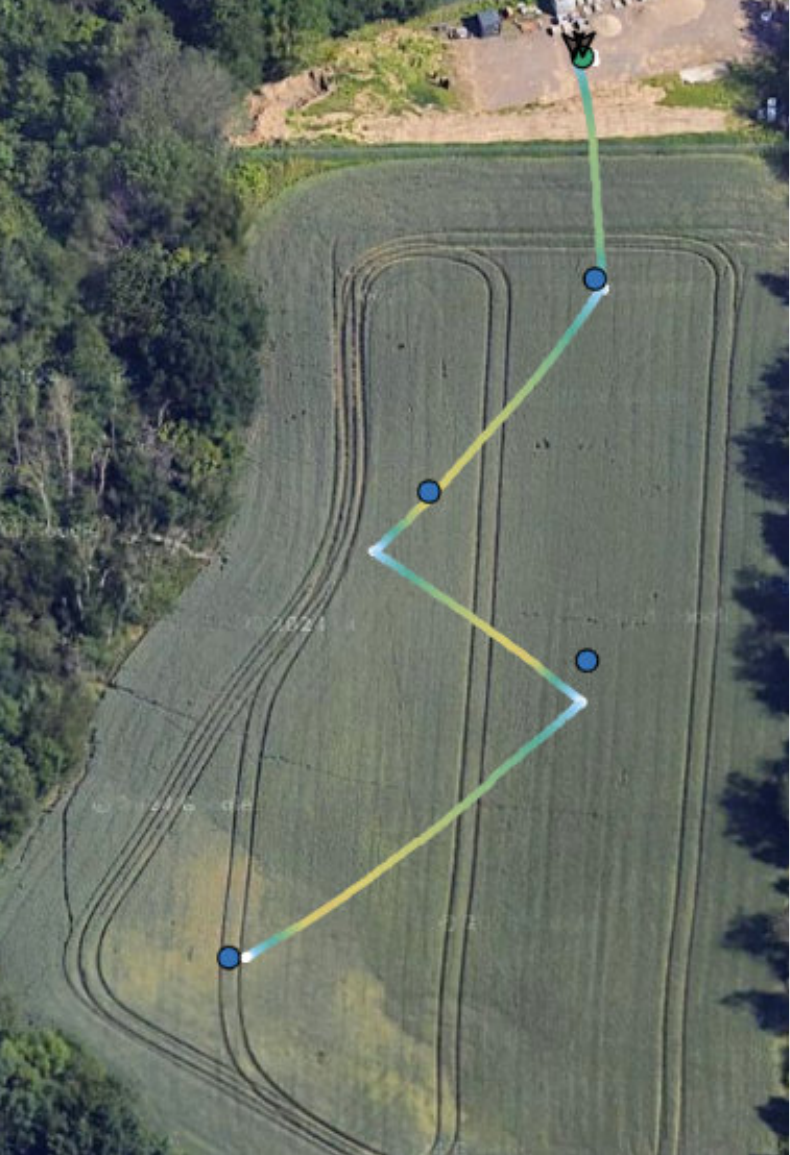}
      \includegraphics[width=0.095\textwidth]{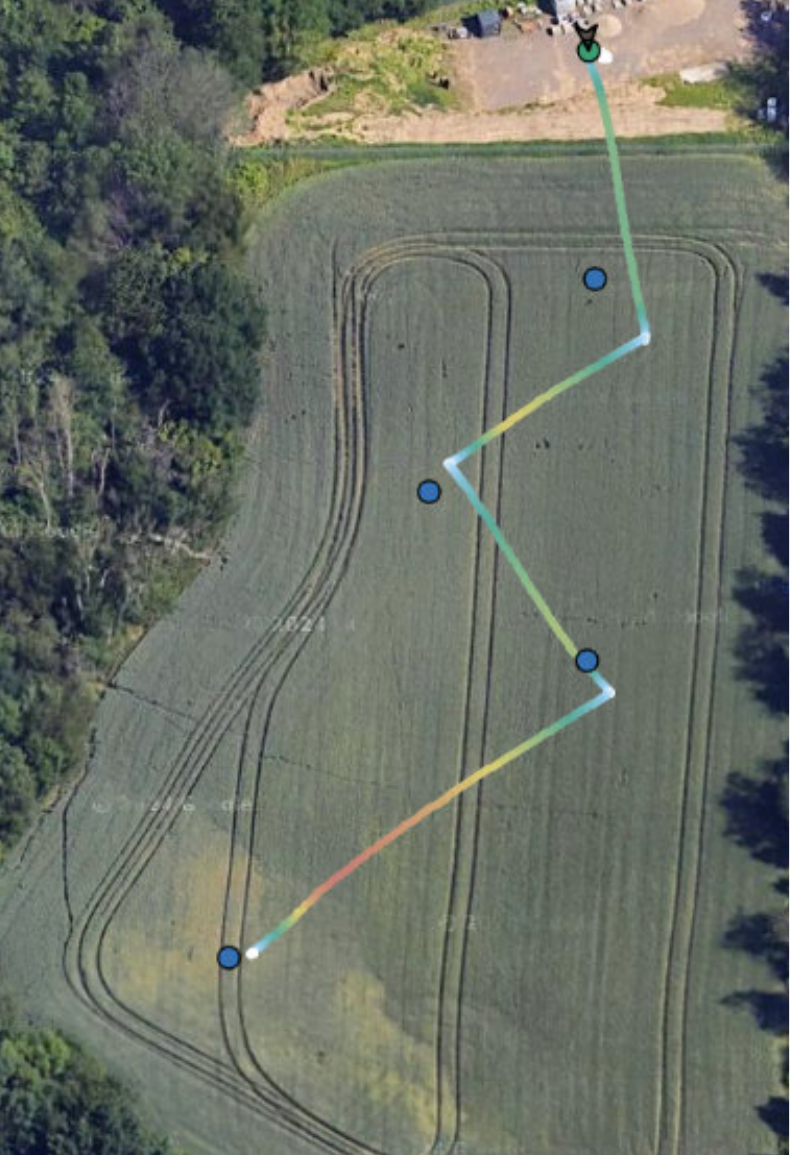}
      \includegraphics[width=0.095\textwidth]{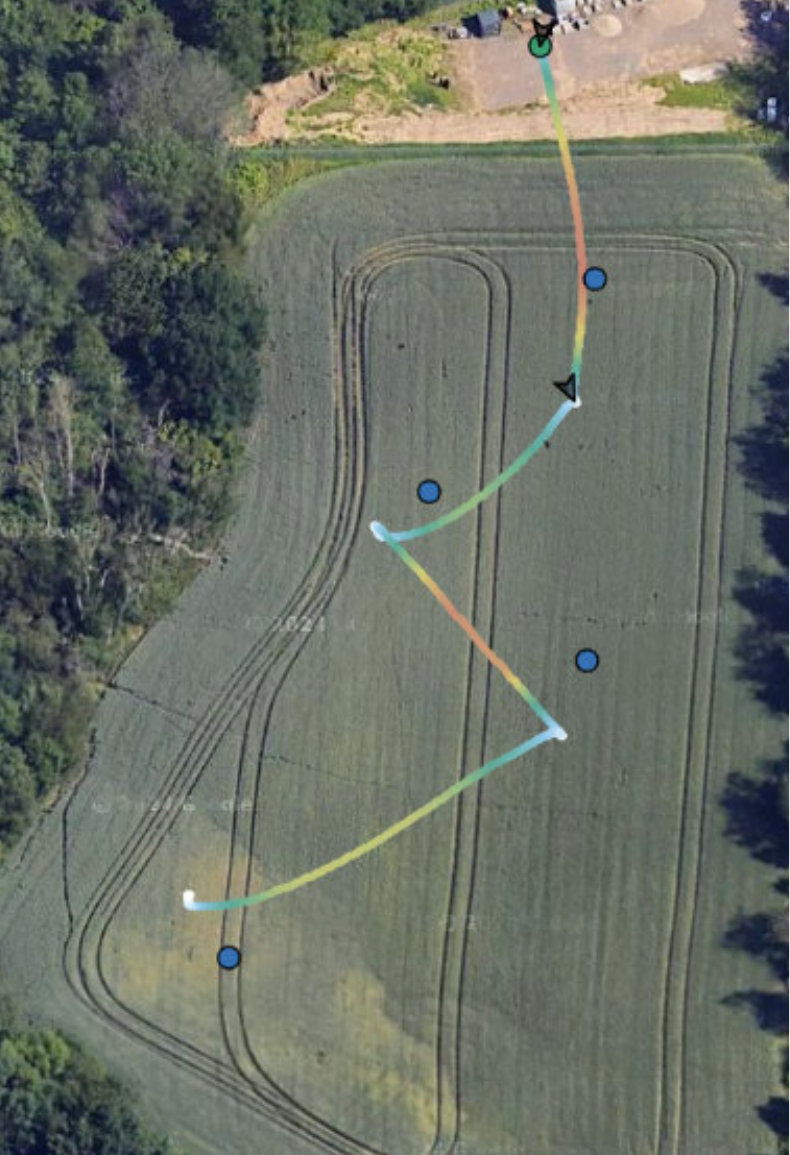}
      \includegraphics[width=0.095\textwidth]{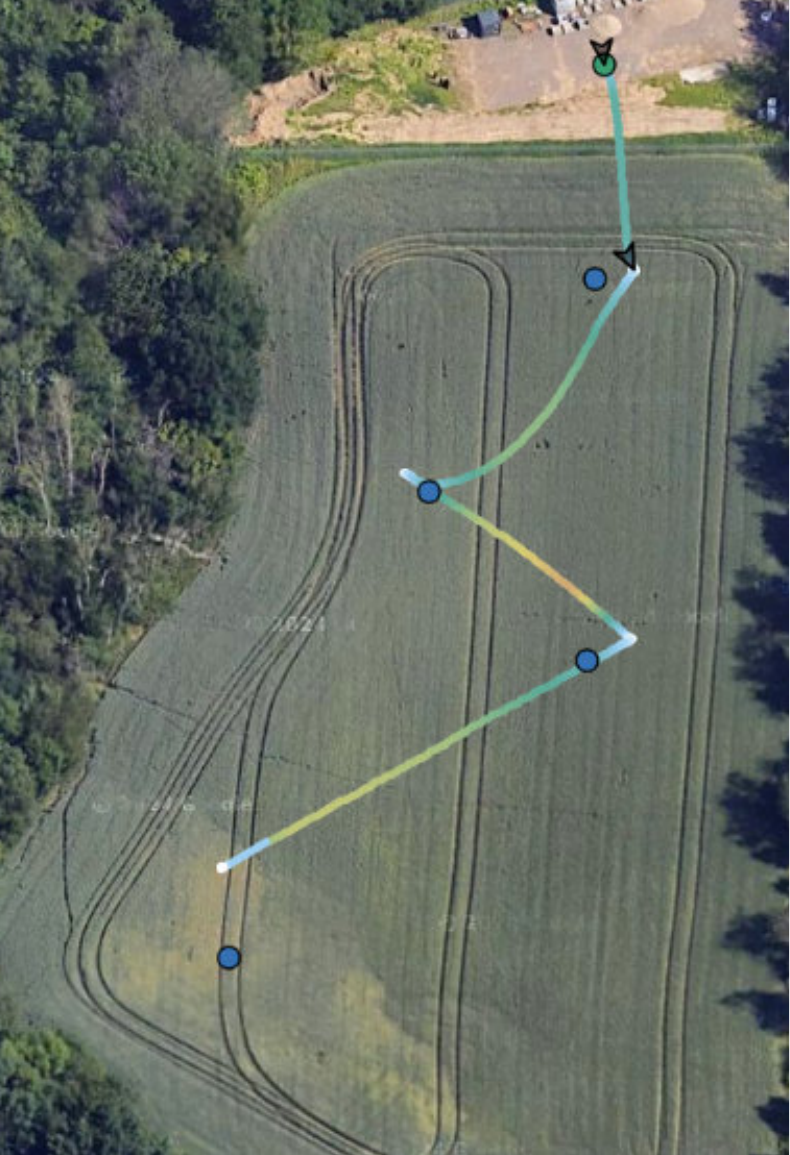}
    \end{minipage}
    \includegraphics[width=\textwidth]{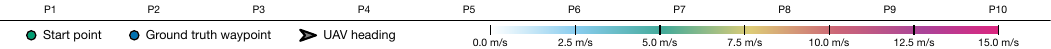}
    \footnotesize Arrows mark heading changes greater than 20°. Path color gradients encode speed.
  \end{minipage}
  \caption{Paths performed for Task~A. Top row: RC. Bottom row: IGUANA 3D map interface.}
  \Description{A collection of 20 pictures showing the UAV path each participant took when conducting Task~A using RC and IGUANA.}
  \label{fig:result-path-task-a}
\end{figure*}

\subsection{Study Design}

\subsubsection{Quantitative Measure}

Participants completed Task~A for this quantitative measure.
To evaluate objective performance, we utilized flight logs generated automatically by the DJI system for each flight.

\paragraph{Task~A: Exploration}

For the exploration task, participants were given a printed map marked with four waypoints.
They were instructed to pilot the \ac{UAV} through each of the four waypoints on the printed map, while avoiding crashing into the terrain below, and to capture a photo of a designated tree at the final waypoint.
Using IGUANA, participants completed this task by entering the 3D map interface, placing waypoint markers, adjusting the virtual camera angle as needed, and then initiating the waypoint mission.
Using the RC, participants manually flew the \ac{UAV} through each waypoint using the dual-stick controller and adjusted the camera gimbal at the final waypoint to take a photo. 

\paragraph{Self Assessment}

We also collected subjective data using post-task and post-test questionnaires.
We used the \ac{SEQ}~\cite{sauroComparisonThreeOnequestion2009} to measure task difficulty, the \ac{NASA-TLX}~\cite{hartDevelopmentNASATLXTask1988} to assess workload, the \ac{SUS}~\cite{brookeSUSQuickDirty1996} for usability, and the \ac{TAM}~\cite{davisRelationshipHCITechnology2007} for user acceptance.

\subsubsection{Qualitative Measure}

Participants completed Task~B as part of the qualitative evaluation.
Unlike Task~A, which focused on objective performance metrics, Task~B was designed to assess subjective impressions, preferences, and perceived ease of use when interacting with the user-centric control mode.

\paragraph{Task~B: Going Home}

This task began once the \ac{UAV} had reached the participant's final waypoint from the exploration task, located on their far right.
Participants were instructed to move the \ac{UAV} from the far right to the center (far in front of them) and then straight toward their position, forming an upside-down capital `L' trajectory.
Using IGUANA, participants used the virtual ball interface in user-centric mode.
In this mode, the task can be completed by pushing the virtual ball to the left until the \ac{UAV} reaches the center, then pulling it towards themselves until the \ac{UAV} returns to the participant.

\paragraph{Interview}

Additionally, a semi-structured interview was conducted after the study to gather qualitative feedback.

\subsection{Participants}

Ten participants (2 female, 8 male; age 20--42, $M=29.8$, $SD=6.3$) took part.
Experience in \ac{XR} (\ac{AR}/\ac{MR}/\ac{VR}), \ac{UAV}, and dual-stick controller was self-rated on a 1--5 scale (1 = beginner, 5 = expert); distributions are shown in~\Cref{fig:participants-demography}.
Each participant received a compensation of \texteuro20 for $\sim$60 minutes of participation.

\subsection{Apparatus}

Apart from the hardware listed in~\Cref{subsec:implementation}, the study setup included: GL.iNet AXT1800 router, iPhone 15 Pro as a mobile hotspot for the router, MacBook Pro for headset casting, Windows 11 laptop running DJI Assistant 2 (Consumer Drones Series) flight simulator~\cite{djiDJIAssistant22018}, and Samsung Galaxy S10 running DJI GO 4~\cite{djiDownloadCenter}.

\subsection{Procedure}

Each session began with an indoor tutorial where participants practiced using each system, with the \ac{UAV} connected to the Windows laptop via a \ac{USB} cable, providing simulated telemetry.
Following the tutorial, the session moved to an open field (approx. 100m x 250m) with ground elevation ranging from 215m to 310m, surrounded by trees.
Participants then used the same setup, with the \ac{UAV} placed on the ground.
An umbrella was used as needed to prevent sunlight from interfering with hand tracking.

We applied counterbalancing to avoid learning effects in our within-subjects design.
Half of the participants started with RC, while the other half started with IGUANA.
After completing each task, participants filled out the \ac{SEQ} and \ac{NASA-TLX} questionnaires.
After completing all tasks for one system, participants filled out \ac{SUS} and \ac{TAM} questionnaires.

Once all questionnaires are completed, the semi-structured interview begins.
We asked open-ended questions about their experience with both our and regular control systems, such as what they liked and disliked about each system, what aspects they found intuitive or confusing, and how they felt during interaction.
Finally, we debriefed the participants and gave them the compensation.

\begin{figure*}[t]
  \centering
  \begin{minipage}{0.8\textwidth}
    \centering
    \begin{minipage}{\textwidth}
      \centering
      \includegraphics[width=0.095\textwidth]{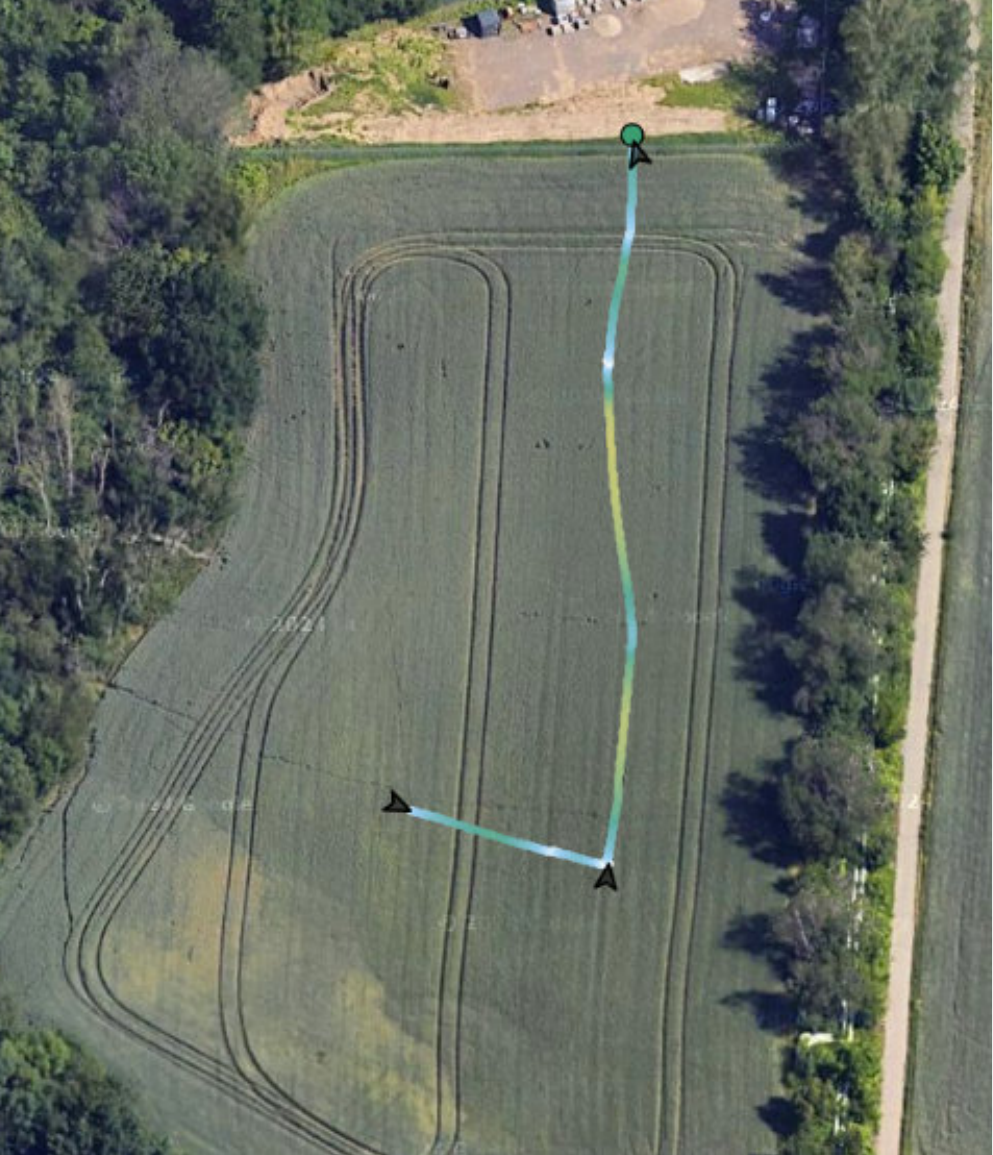}
      \includegraphics[width=0.095\textwidth]{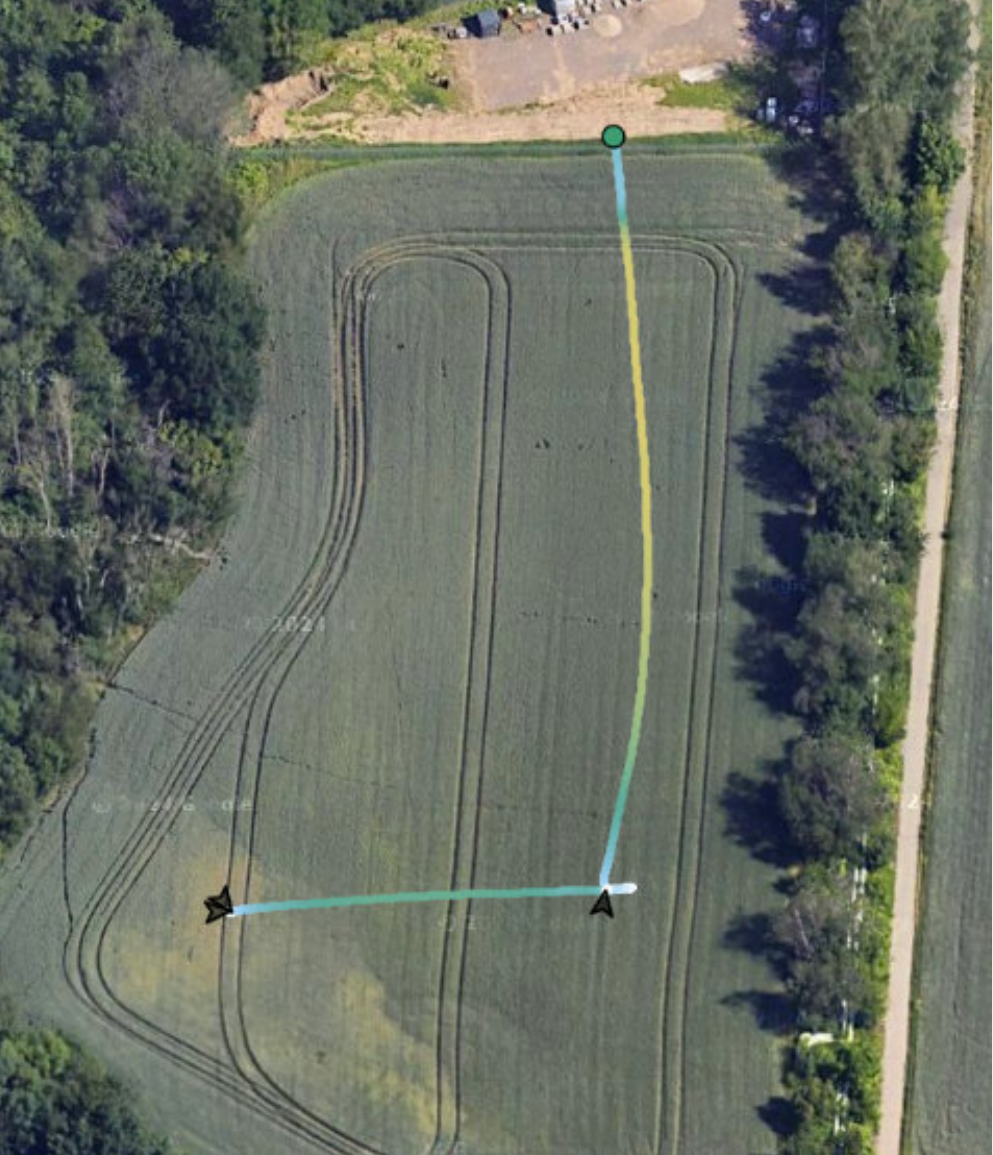}
      \includegraphics[width=0.095\textwidth]{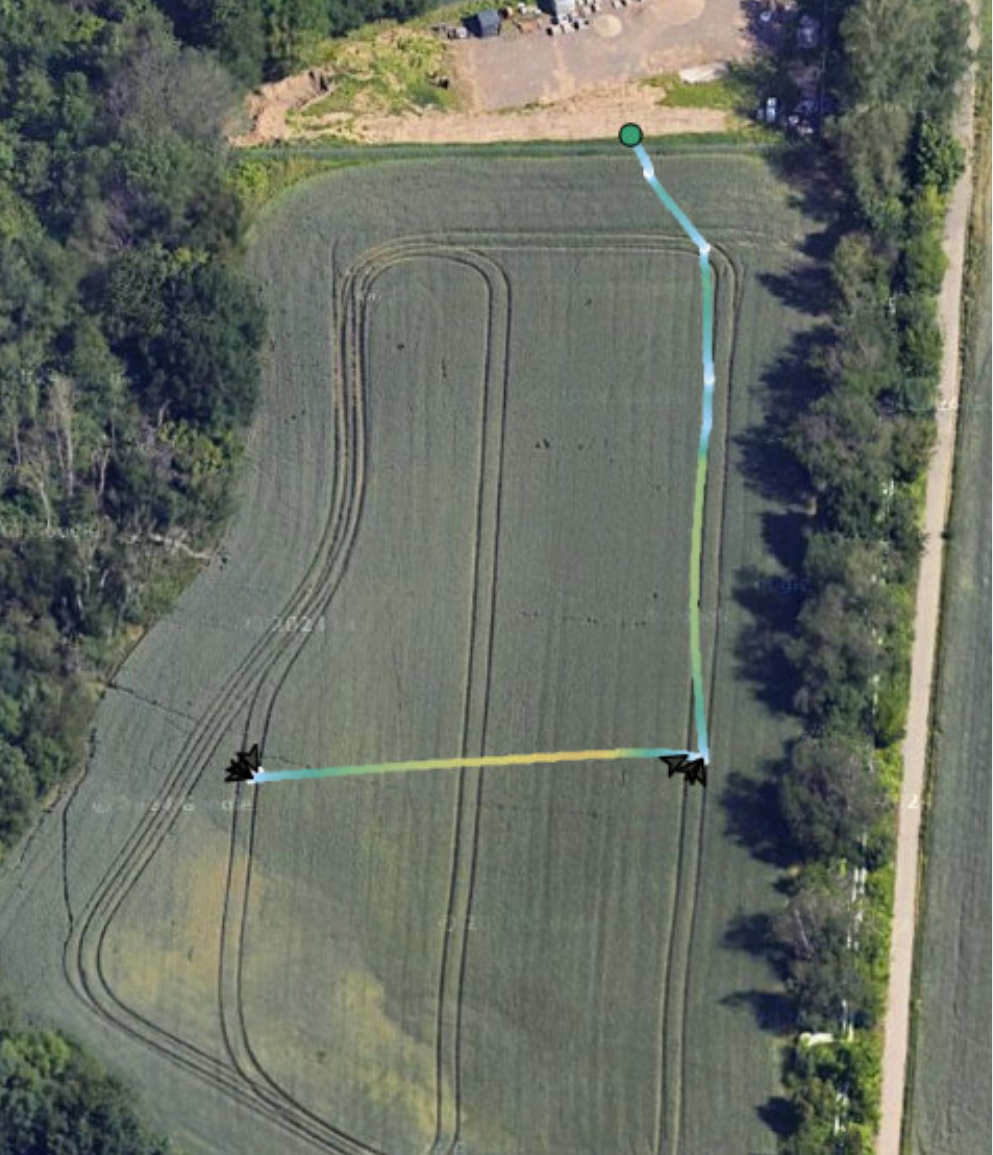}
      \includegraphics[width=0.095\textwidth]{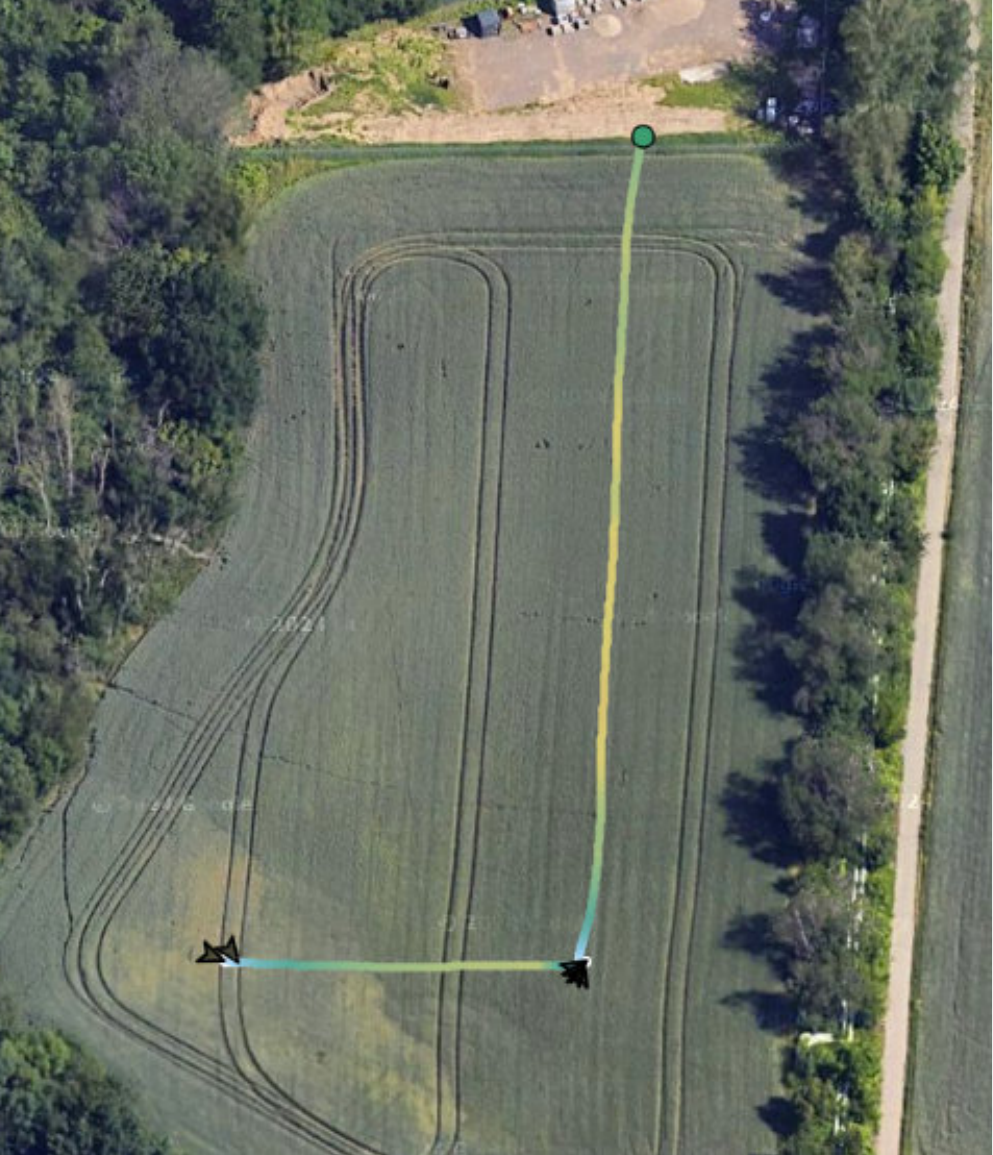}
      \includegraphics[width=0.095\textwidth]{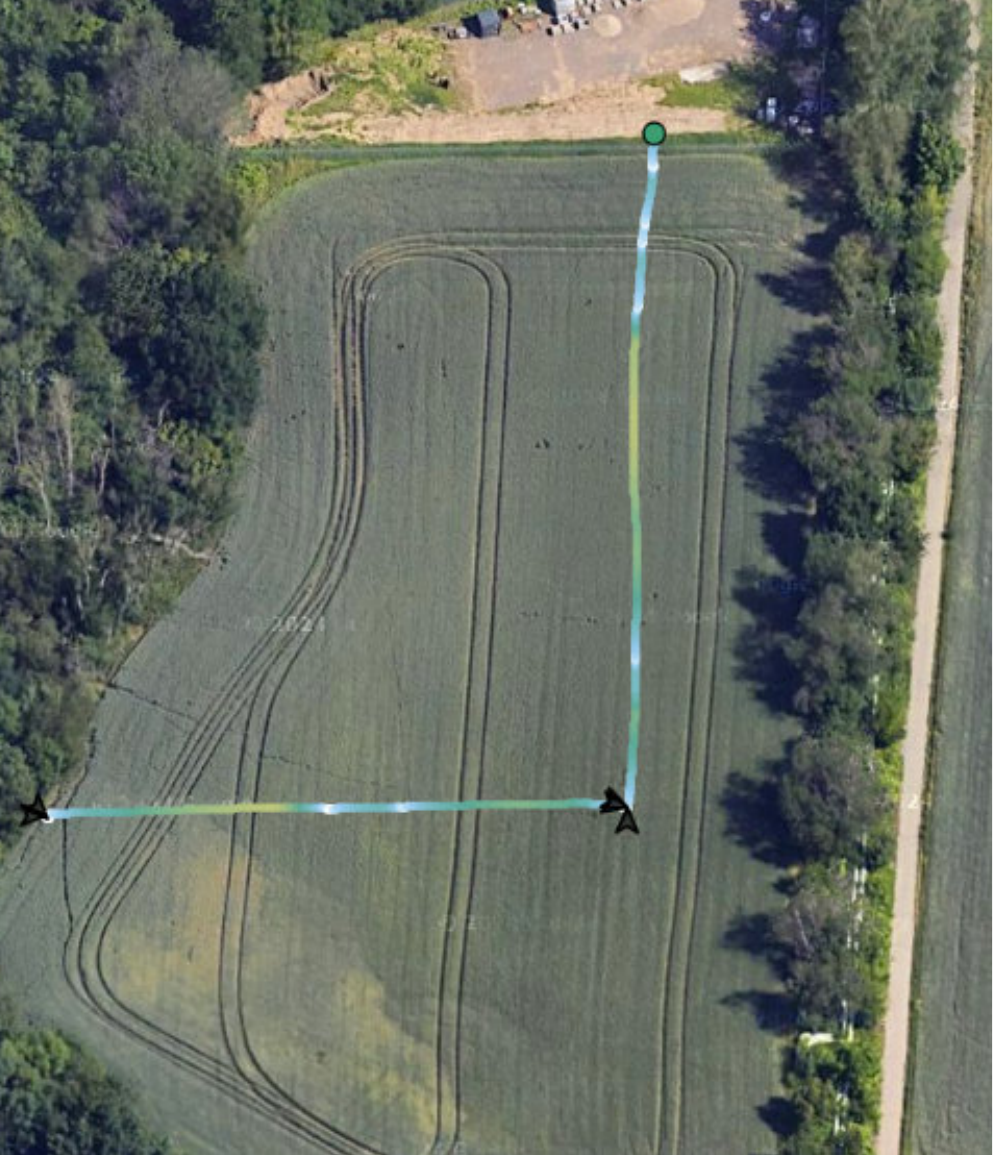}
      \includegraphics[width=0.095\textwidth]{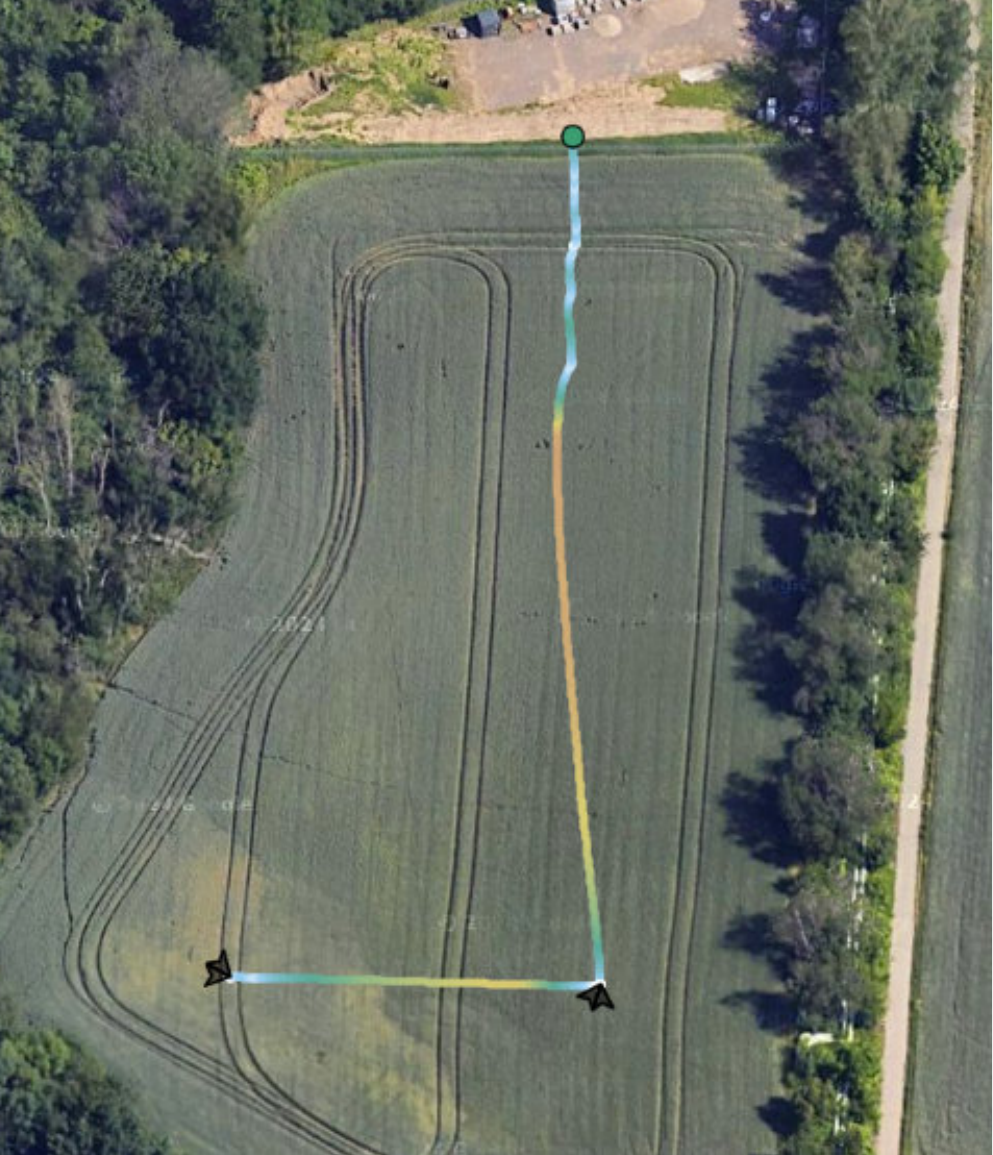}
      \includegraphics[width=0.095\textwidth]{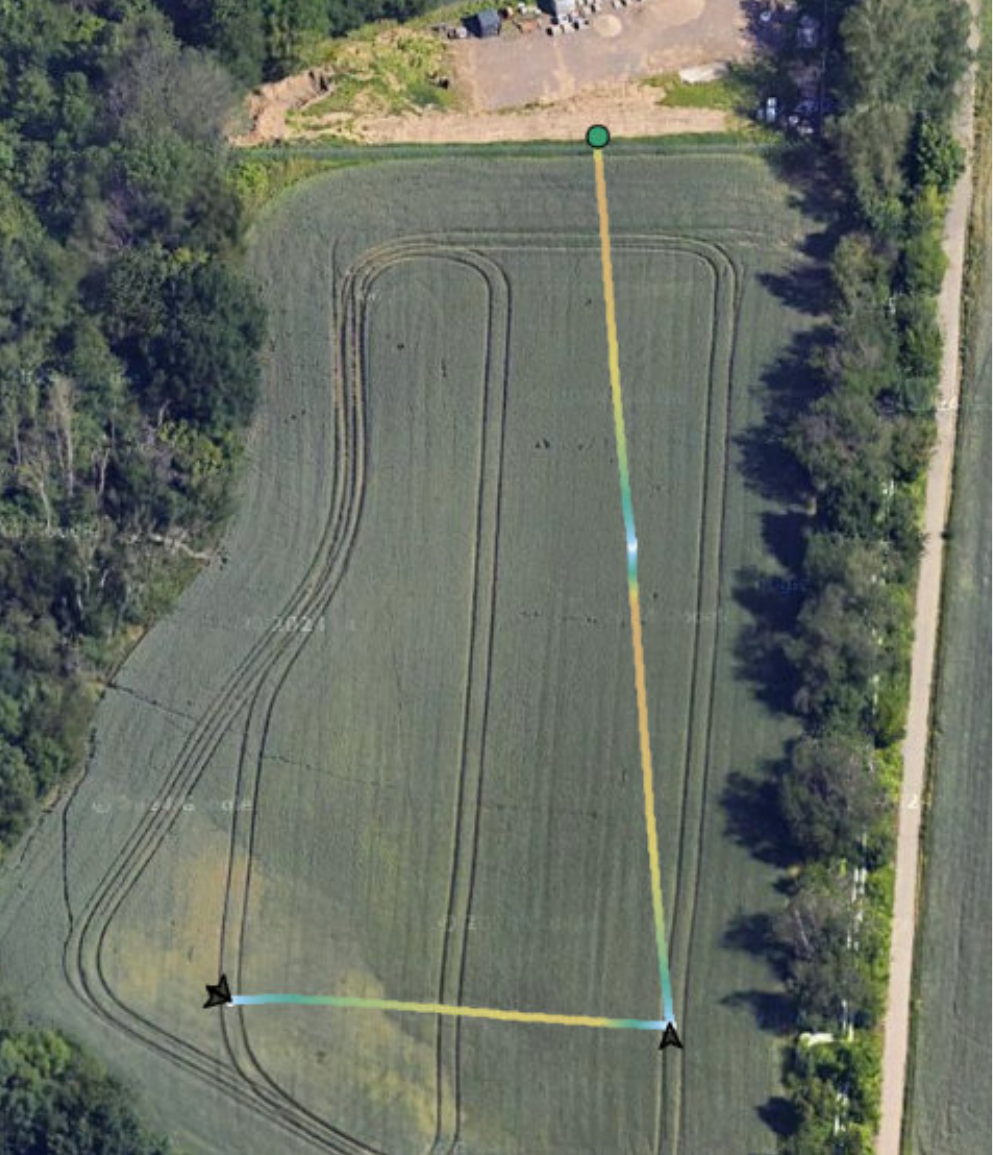}
      \includegraphics[width=0.095\textwidth]{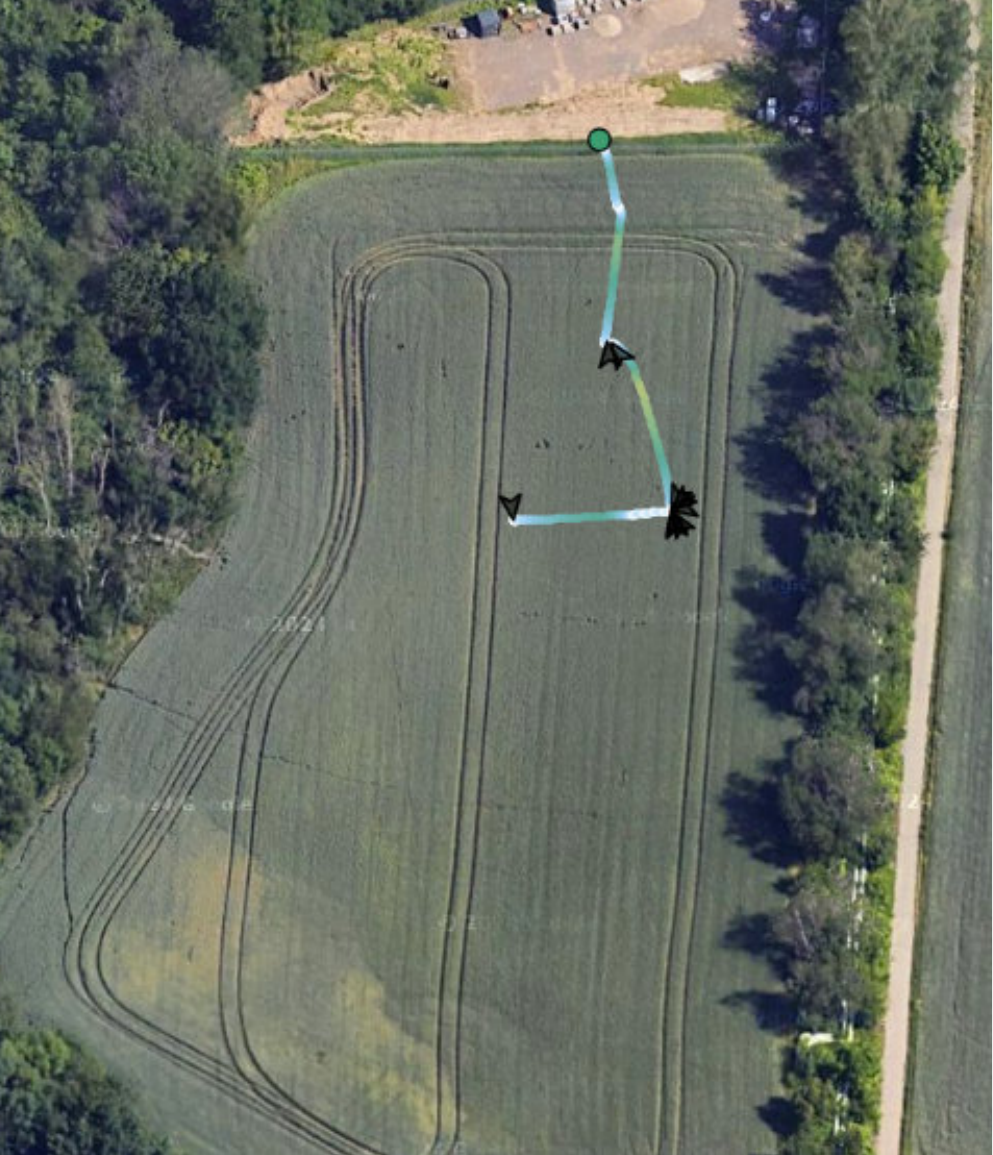}
      \includegraphics[width=0.095\textwidth]{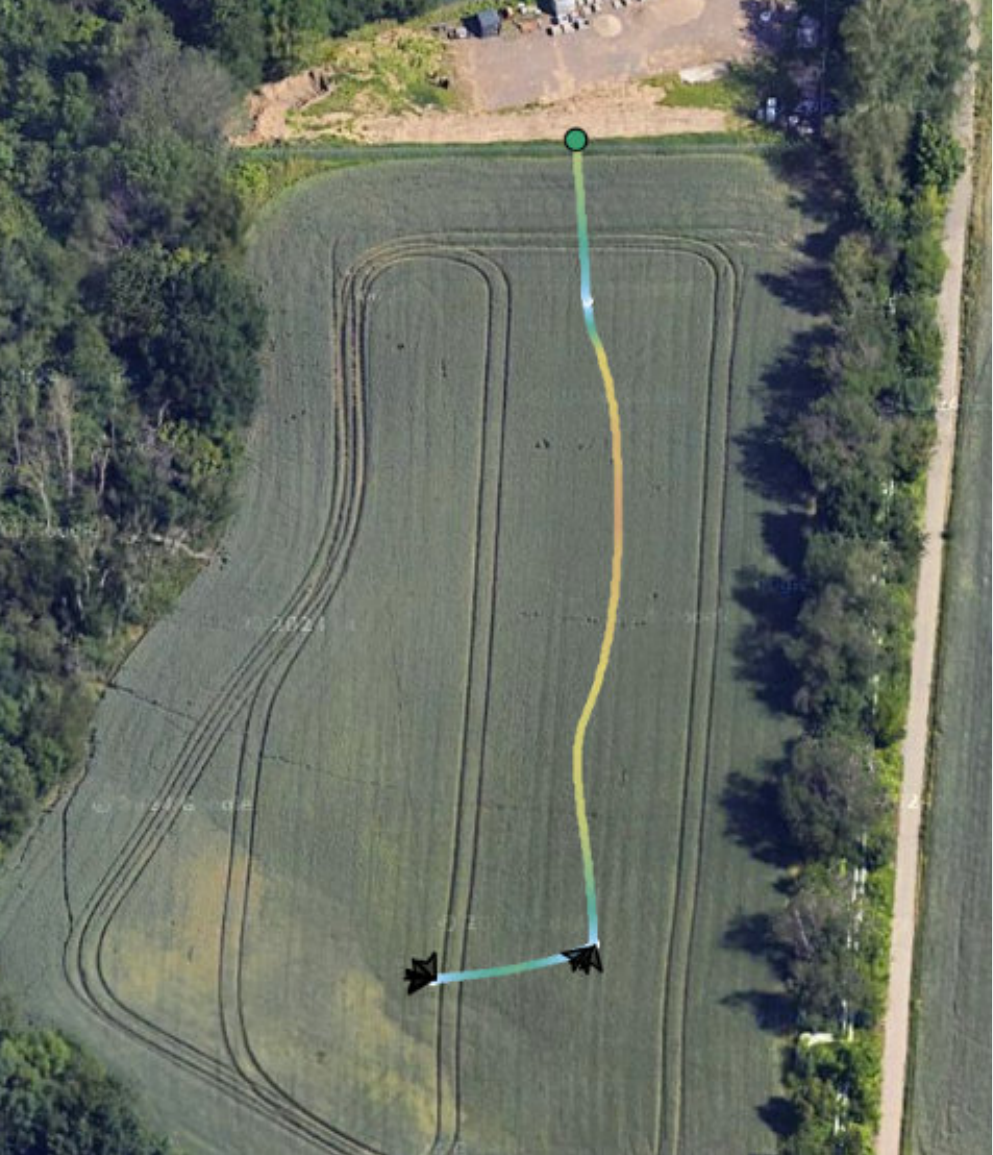}
      \includegraphics[width=0.095\textwidth]{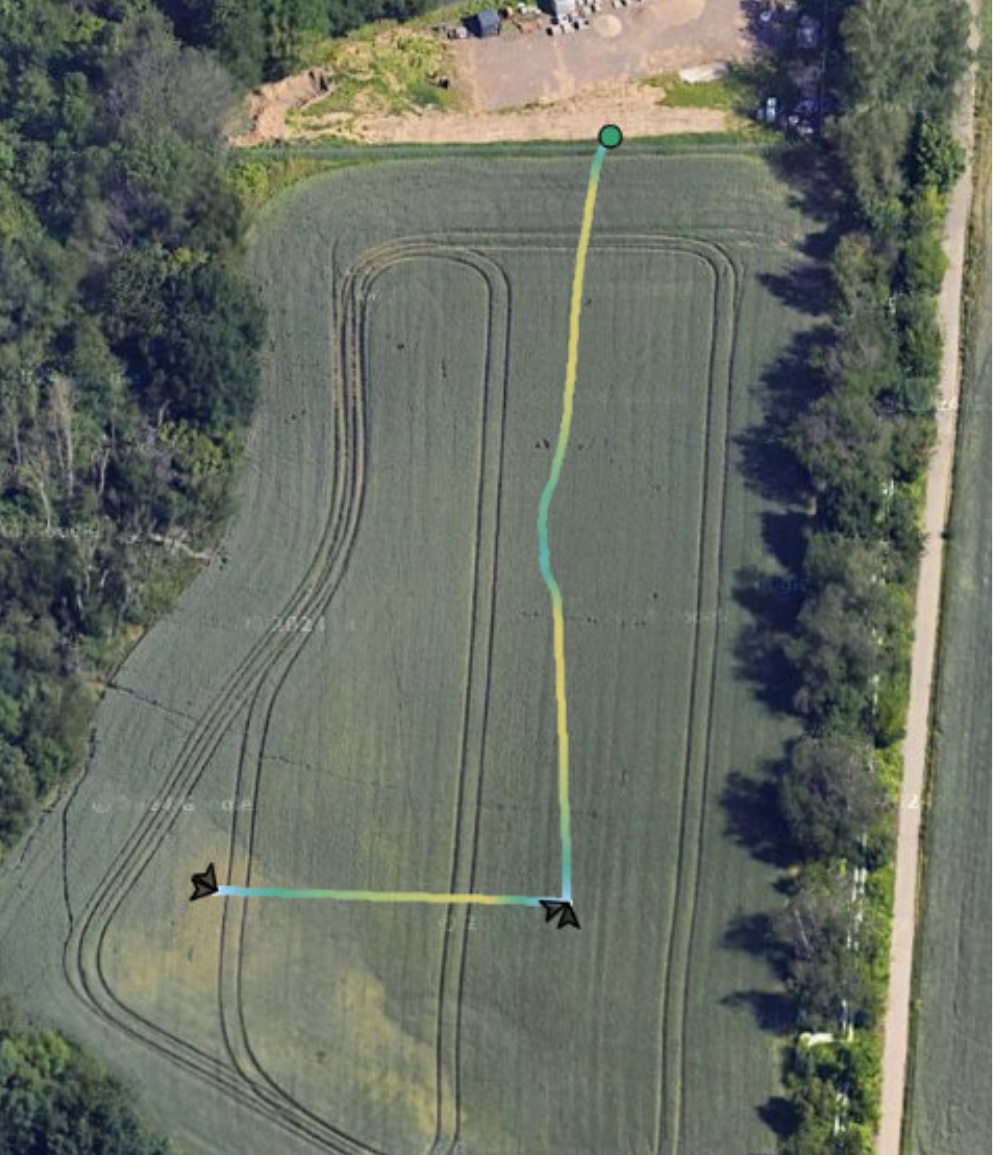}
    \end{minipage}
    \begin{minipage}{\textwidth}
      \centering
      \includegraphics[width=0.095\textwidth]{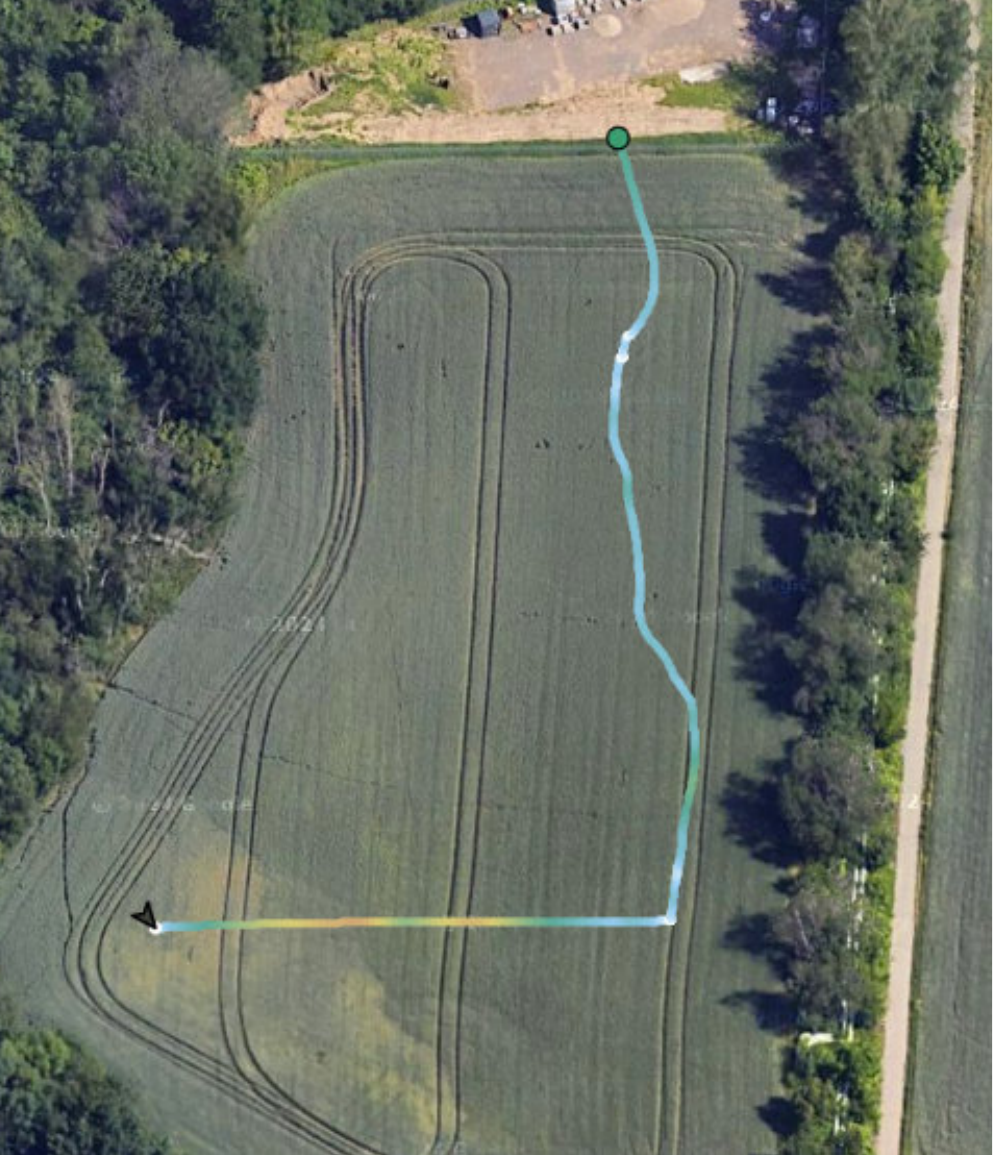}
      \includegraphics[width=0.095\textwidth]{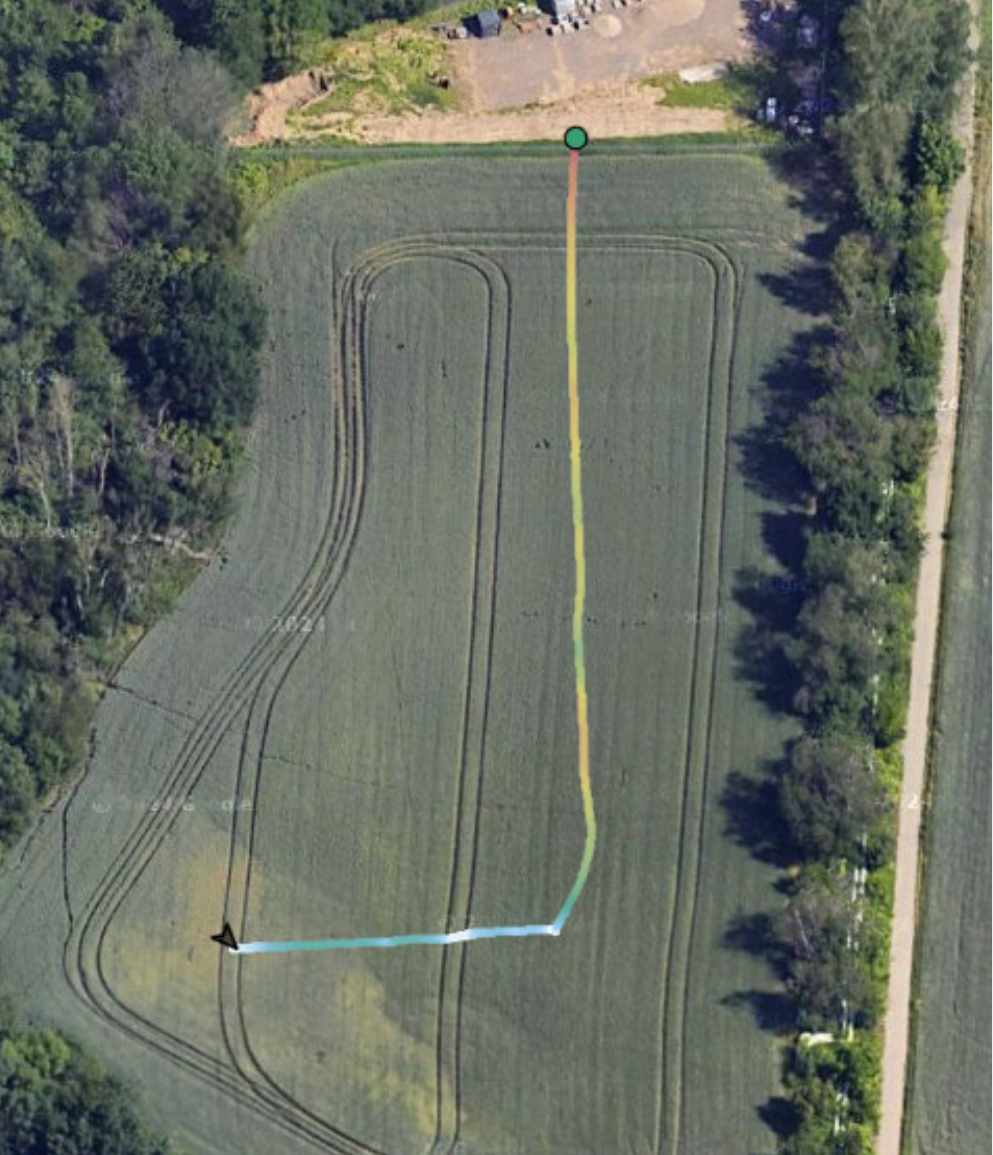}
      \includegraphics[width=0.095\textwidth]{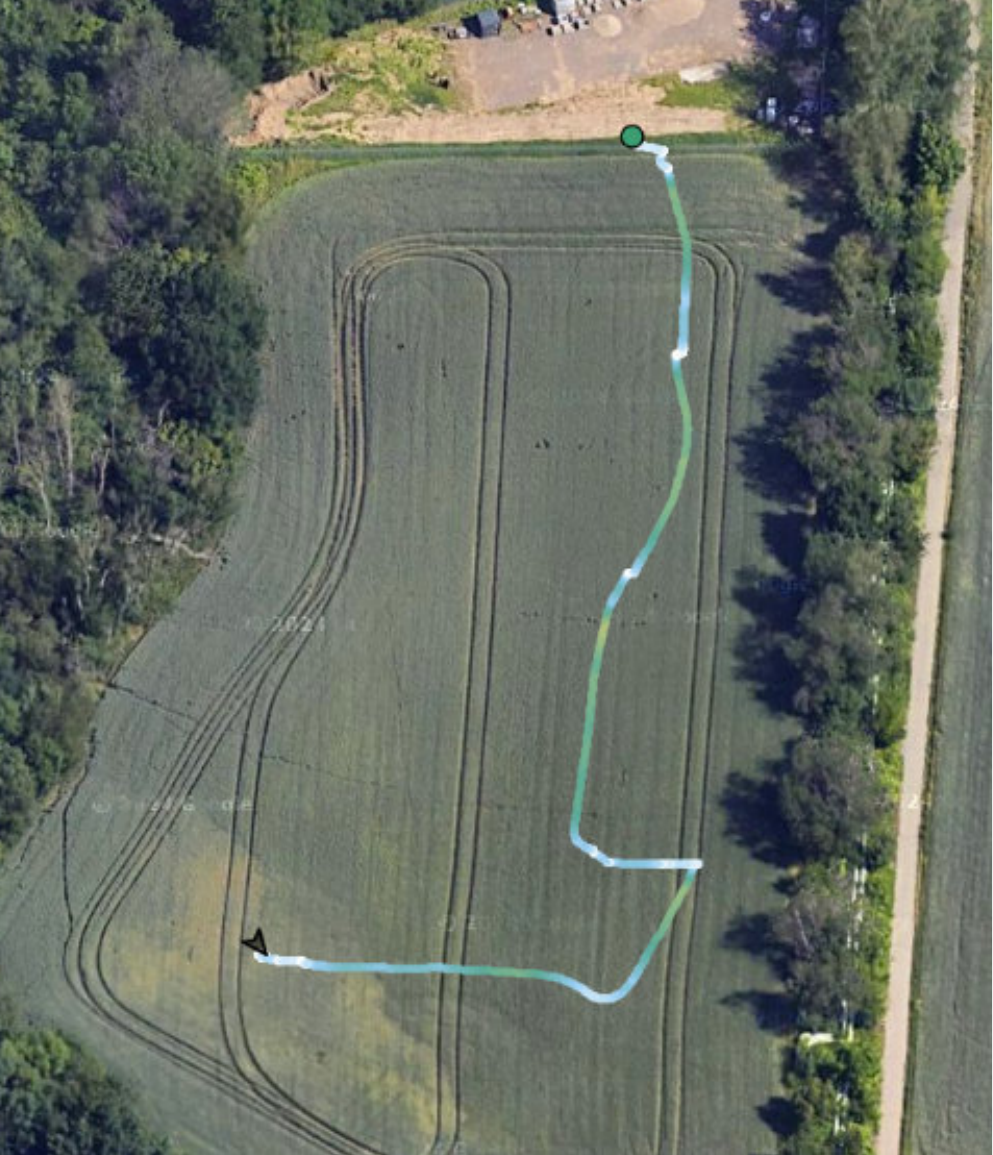}
      \includegraphics[width=0.095\textwidth]{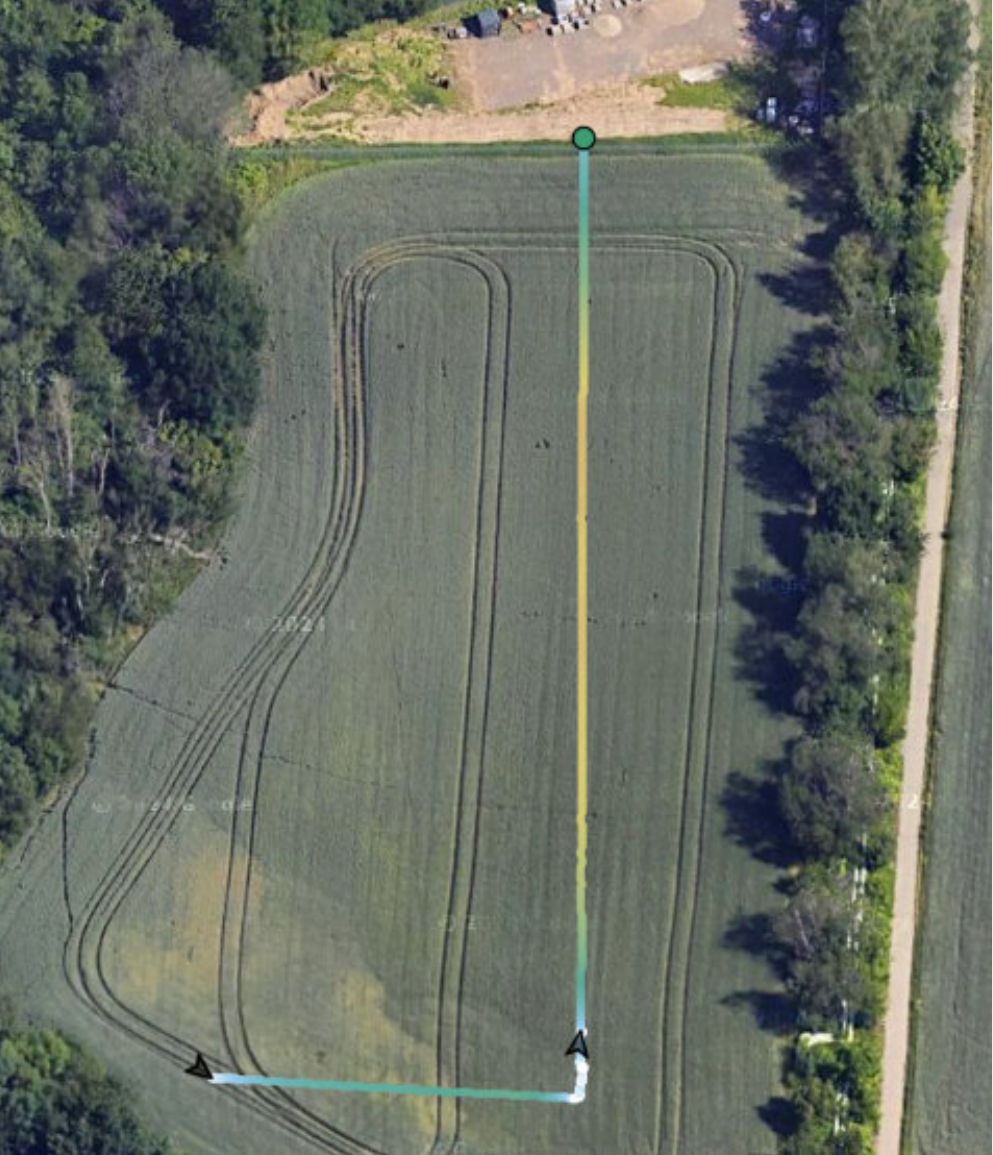}
      \includegraphics[width=0.095\textwidth]{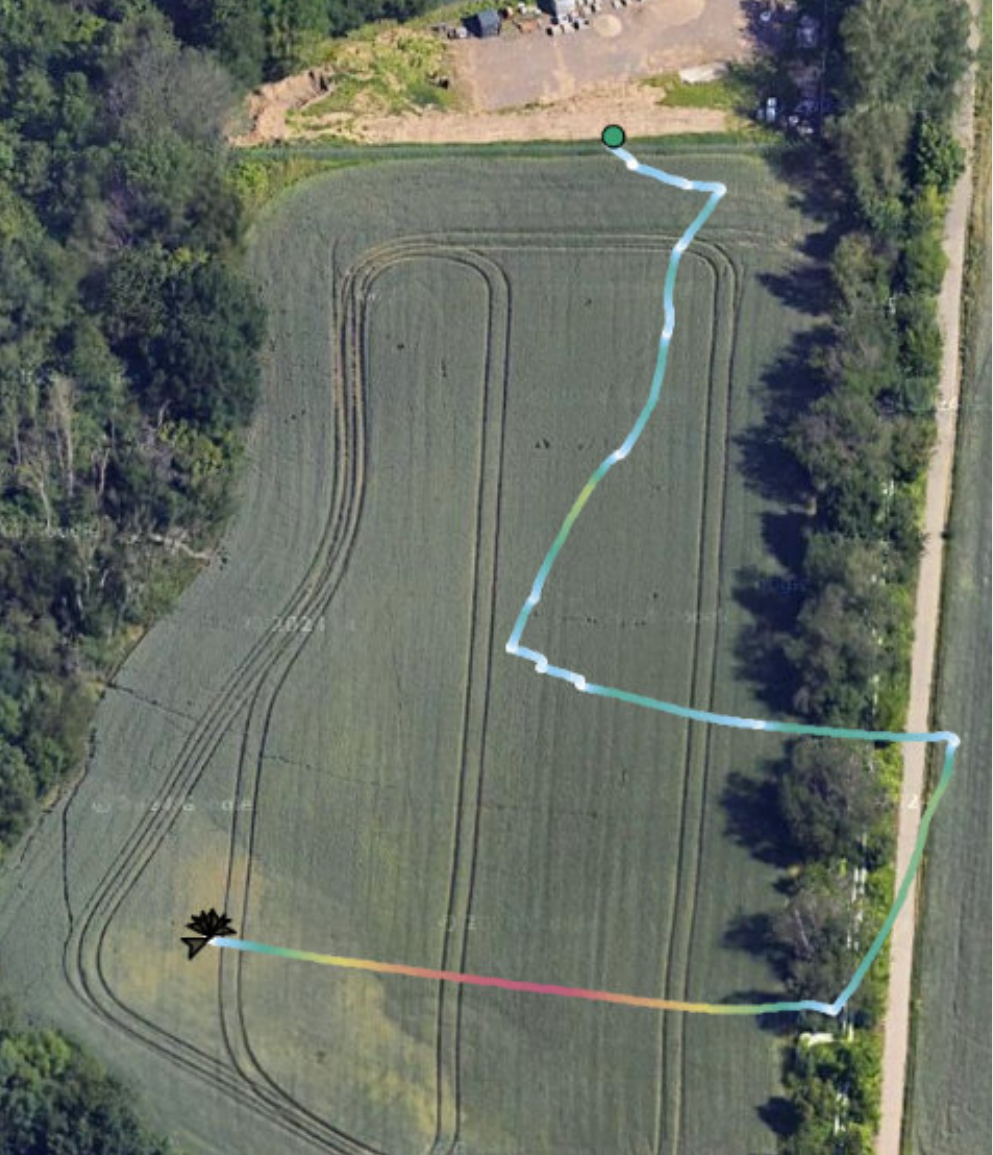}
      \includegraphics[width=0.095\textwidth]{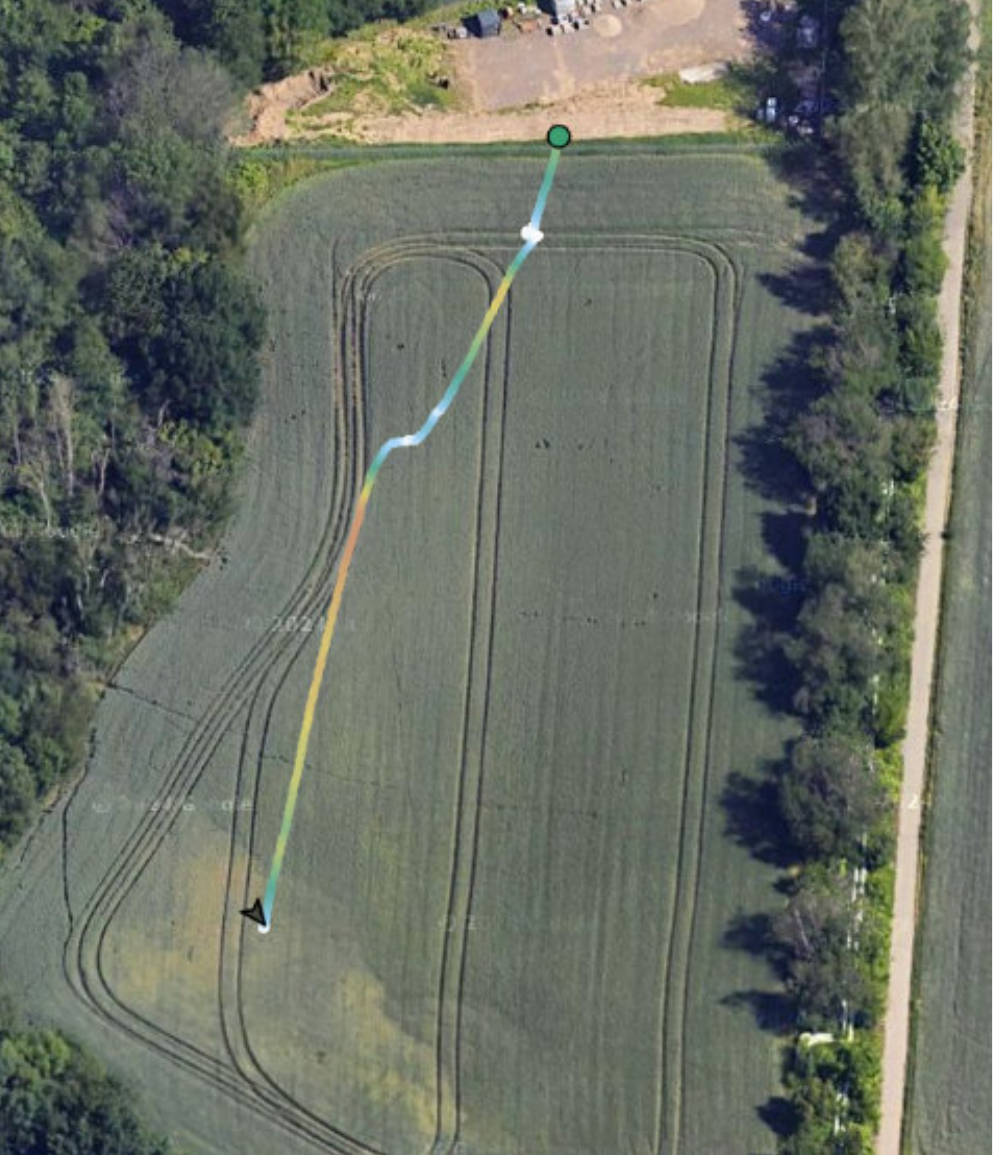}
      \includegraphics[width=0.095\textwidth]{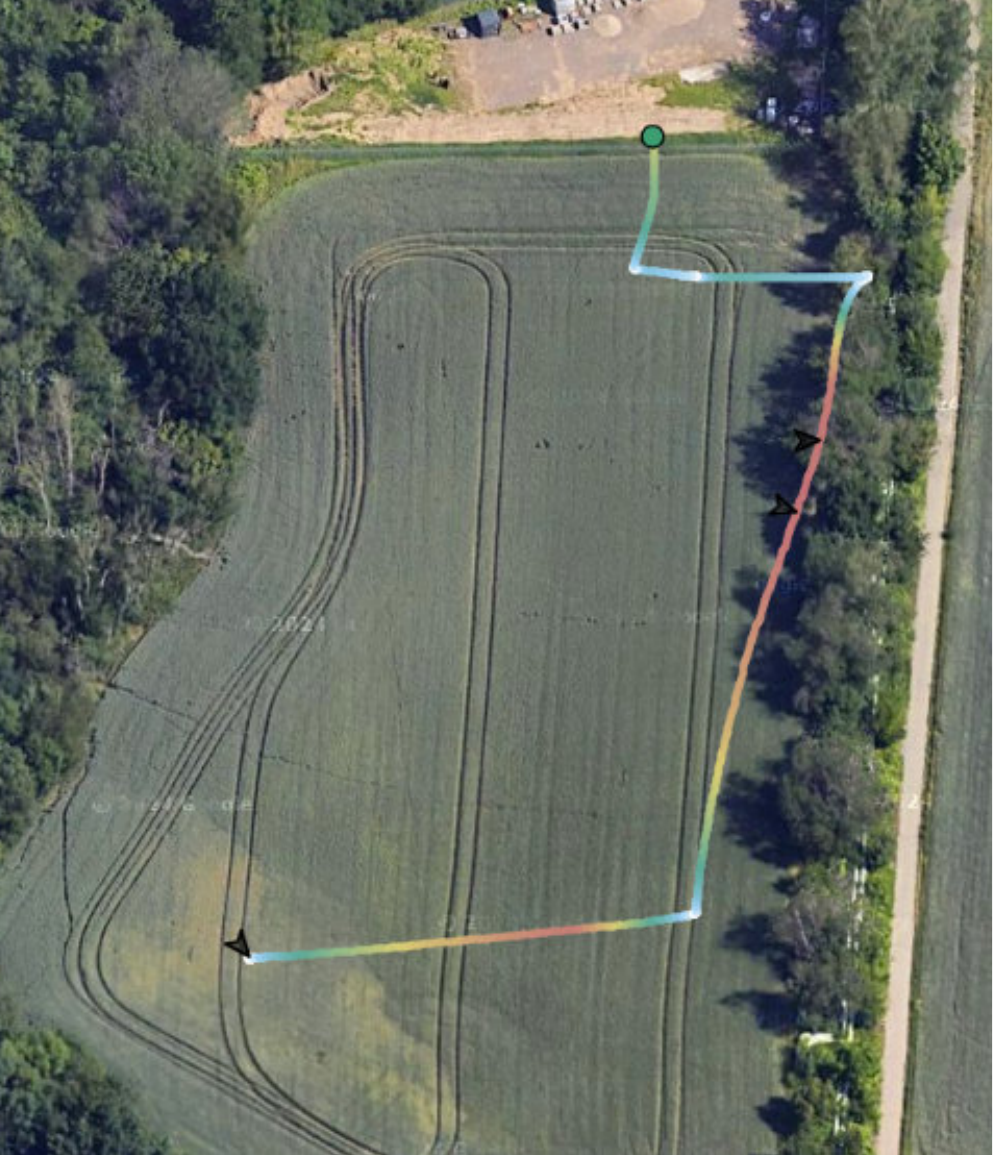}
      \includegraphics[width=0.095\textwidth]{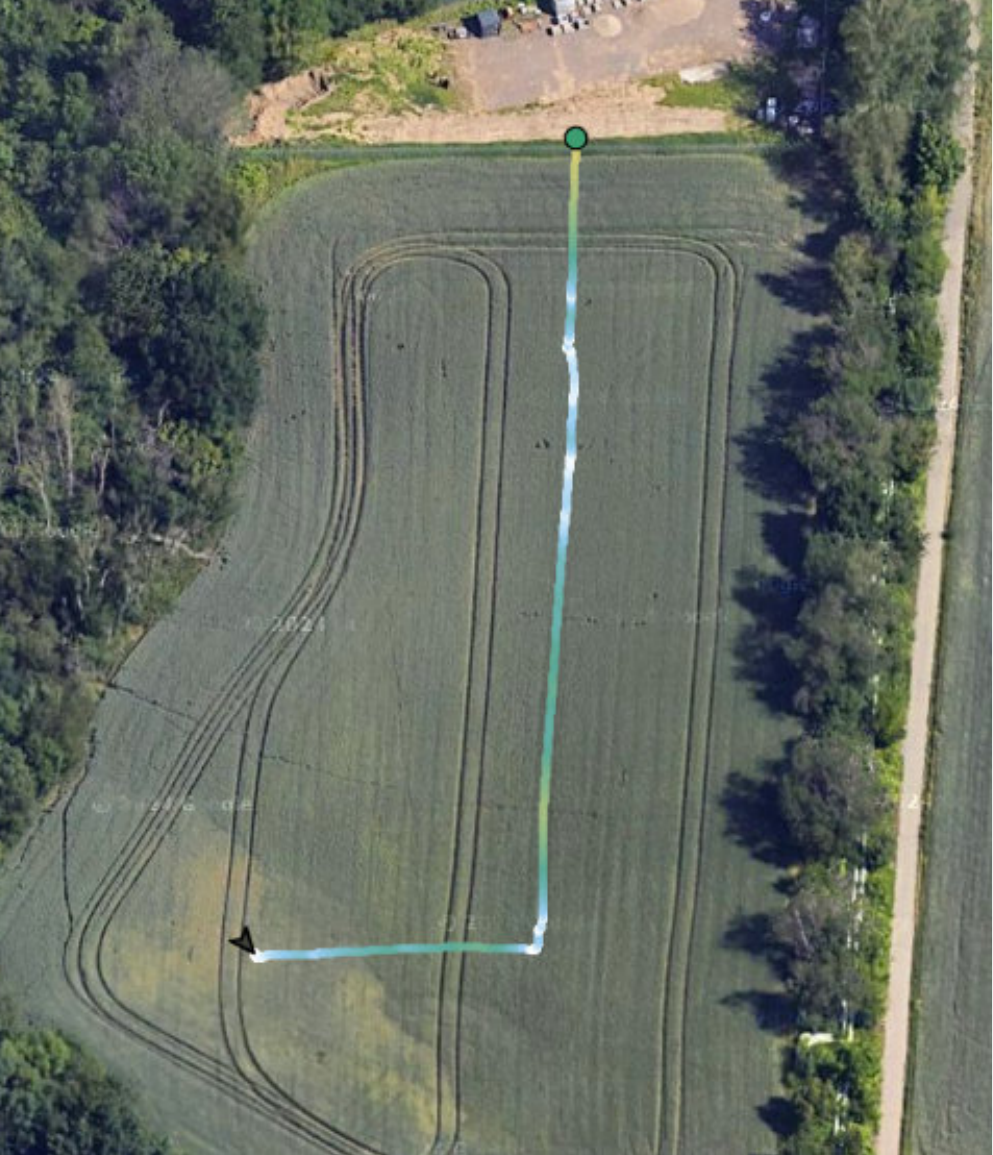}
      \includegraphics[width=0.095\textwidth]{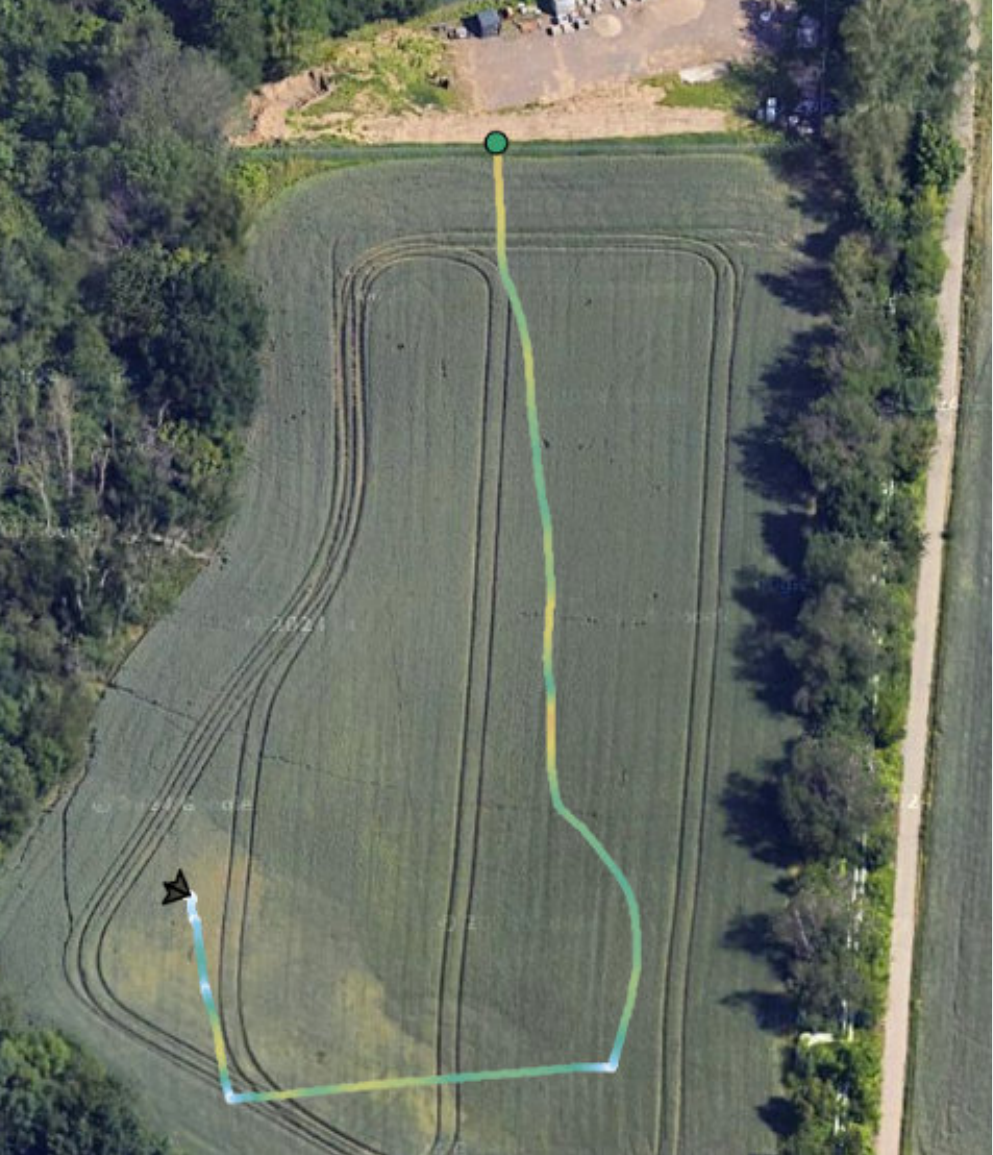}
      \includegraphics[width=0.095\textwidth]{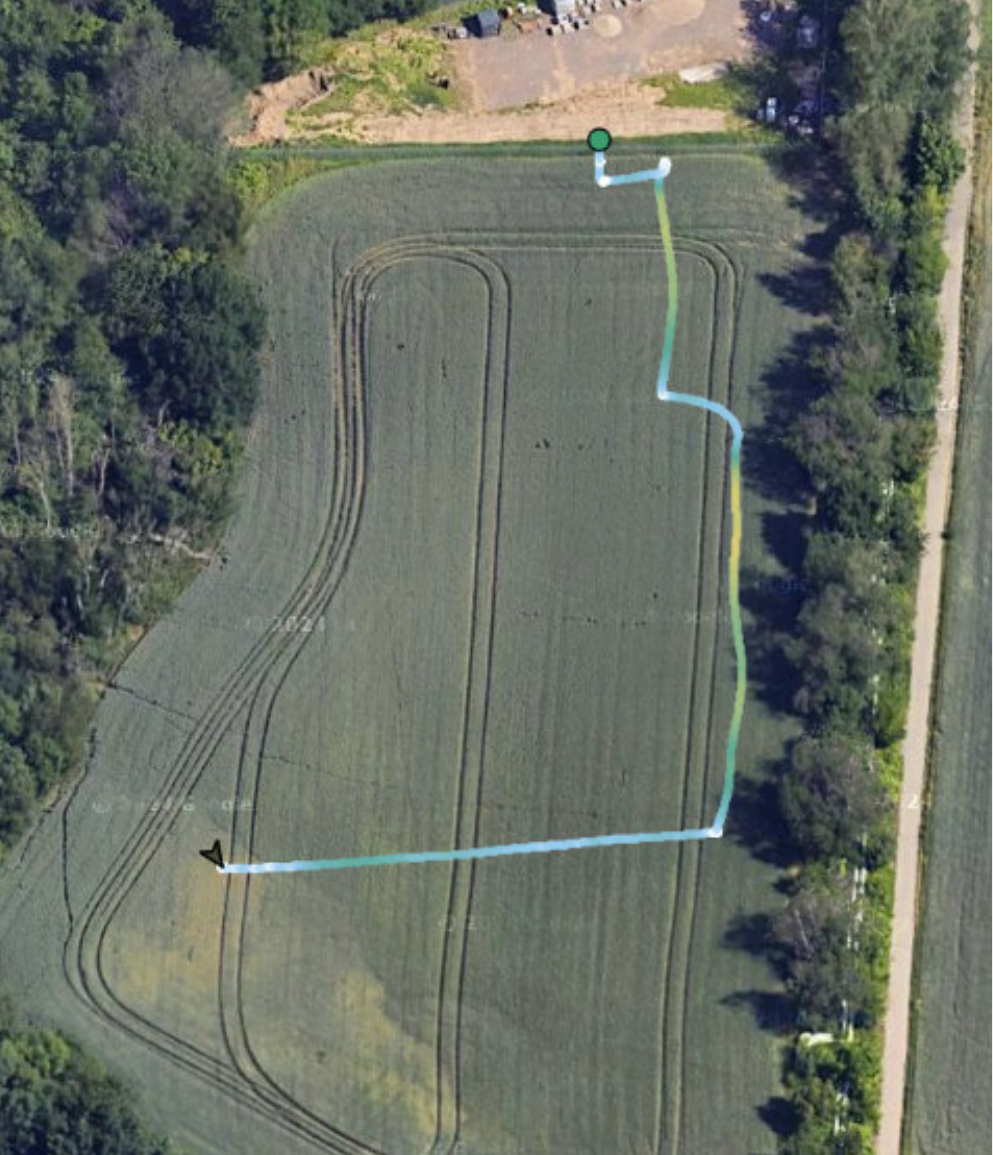}
    \end{minipage}
    \includegraphics[width=\textwidth]{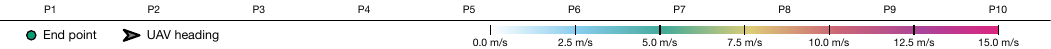}
    \footnotesize Arrows mark heading changes greater than 20°. Path color gradients encode speed.
  \end{minipage}
  \caption{Paths performed for Task~B. Top row: RC. Bottom row: IGUANA virtual ball interface (user-centric).}
  \Description{A collection of 20 pictures showing the UAV path each participant took when conducting Task~B using RC and IGUANA.}
  \label{fig:result-path-task-b}
\end{figure*}

\subsection{Data Analysis}

For each participant, we collected two flight logs: one from their IGUANA flight and one from their RC flight.
These logs contain detailed telemetry data, including position, velocity, and control inputs.
From this data, we derived a performance score by comparing the actual flight trajectory to the predefined waypoints and calculated the task completion time.
Photos taken were also collected to verify task completion.

\paragraph{Task~A: Exploration}

The performance score was calculated using ~\Cref{eq:score_formula}.
\begin{equation}
S = \dfrac{
  \overline{D}_{\mathrm{wp}} \; + \;
  \begin{cases}
  \left(1 - \dfrac{T - T_{\min}}{T_{\max} - T_{\min}}\right) + P & \text{if } D_{\mathrm{wp}_4} < \delta, \\
  0 & \text{otherwise}
\end{cases}
}{3},
\label{eq:score_formula}
\end{equation}
where $\overline{D}_{\mathrm{wp}}$ is the average closest distance score,
$D_{\mathrm{wp}_4}$ is the closest distance score to the final waypoint,
$T$ is the participant's completion time, $T_{\min}$ and $\ T_{\max}$ are the minimum and maximum completion times across all participants,
and $P$ is a binary photo indicator ($1$ if a photo was taken, $0$ otherwise).
A waypoint was considered passed if the \ac{UAV}'s trajectory came within a predefined distance threshold ($\delta=10$ m).
Completion time and photo were only factored into the score if the final (farthest) waypoint was successfully reached, ensuring that the object was captured from a reasonable distance to be considered valid.

\section{Results}
\label{sec:results}

To support our analysis, we developed a custom JavaScript tool that parses the flight logs and visualizes the \ac{UAV}'s trajectory on a map, along with relevant telemetry data (e.g., speed, heading).
The resulting visualizations are shown in~\Cref{fig:result-path-task-a} and \ref{fig:result-path-task-b}, illustrating that for Task~A, all participants produced similar trajectories with the 3D map interface (see~\cref{fig:result-path-task-a}, bottom row), in contrast to the more varied paths observed with RC (see~\cref{fig:result-path-task-a}, top row).

\begin{figure*}[t]
  \centering
  \includegraphics[width=\textwidth]{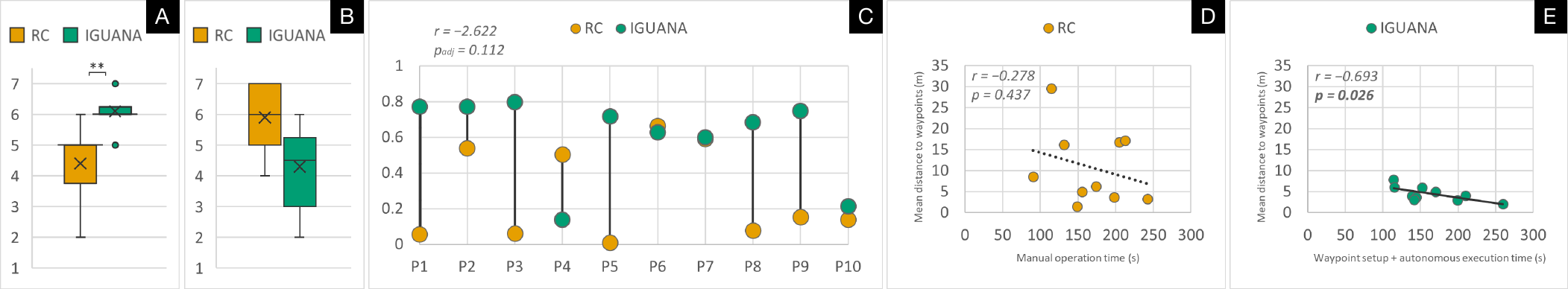}
  \footnotesize Significance: ** $p_{\text{adj}}<0.01$. Trend lines: dashed = not significant, solid = significant ($p<0.05$).
  \caption{Boxplots of SEQ scores for (A) Task~A and (B) Task~B. (C) Dumbbell plot comparing performance scores, each line connects a participant's paired scores across RC and IGUANA. Scatter plots showing the relationship between completion time and accuracy (mean distance to waypoints) for (D) RC and (E) IGUANA.}
  \Description{Boxplots of Single Ease Question scores for Task~A and Task~B, dumbbell plot comparing participants' paired performance scores across RC and IGUANA, and scatter plots showing the relationship between completion time and accuracy when using RC and IGUANA.}
  \label{fig:result-seq-time-distance}
\end{figure*}

\subsection{Quantitative Results}

To prepare the data for statistical analysis, we first assessed the normality of the paired differences for all comparisons using Shapiro-Wilk tests and Q-Q plots.
The assumption of normality was met for all variables ($p=.061$--$.978$) except for Satisfaction in the \ac{TAM} ($p=.003$), Physical Demand in the \ac{NASA-TLX} for Task~A ($p=.001$), and Temporal Demand in the \ac{NASA-TLX} for Task~B ($p=.034$).
Therefore, Wilcoxon signed-rank tests were used for these three comparisons, and paired t-tests were used for the rest.
To control the family-wise error rate, we applied Holm-Bonferroni corrections ($\alpha=.05$) within three predefined families of tests:
\ac{SUS} and \ac{TAM} ($m=6$),
Task~A measures including \ac{SEQ}, \ac{NASA-TLX}, completion time, average closest distance to waypoints, and performance score ($m=10$), and
Task~B measures including \ac{SEQ} and \ac{NASA-TLX} ($m=7$).
All analyses were conducted in JASP~\cite{JASP2025}.

\begin{figure}[t]
  \centering
  \includegraphics[width=\columnwidth]{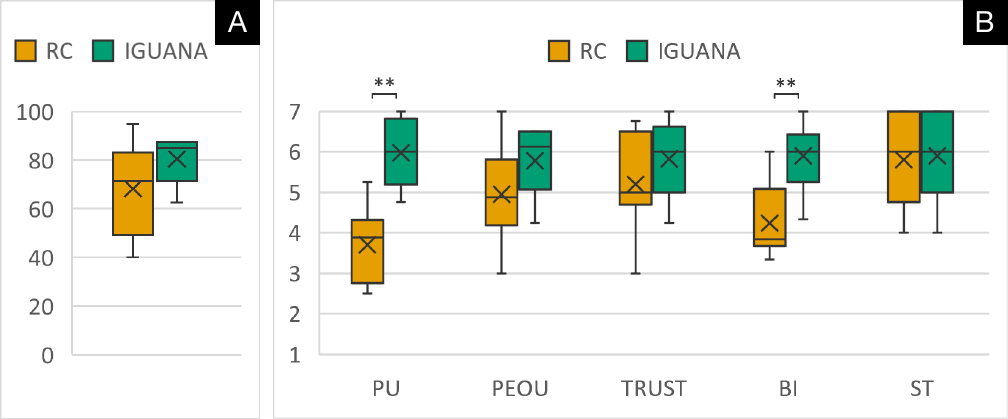}
  \footnotesize PU = Perceived Usefulness, PEOU = Perceived Ease of Use, TRUST = Trust in the System, BI = Behavioral Intention to Use, ST = System Transparency.\\Significance: ** $p_{\text{adj}}<0.01$.
  \caption{Boxplots of (A) SUS scores and (B) TAM scores.}
  \Description{Boxplots of System Usability Scale and Technology Acceptance Model for RC and IGUANA.}
  \label{fig:result-sus-tam}
\end{figure}

\ac{SUS} scores were generally higher for IGUANA ($M=80.5$, $SD=2.98$) than for RC ($M=68.25$, $SD=5.89$) (see~\cref{fig:result-sus-tam}-A).
A paired-t test results in a non-significant difference ($t(9)=2.293$, $p_{\text{adj}}=.192$), indicating usability was comparable to RC.

Paired t-tests on the \ac{TAM} components revealed that IGUANA received significantly higher ratings than RC for Perceived Usefulness ($t(9)=5.609$, $\boldsymbol{p_{\text{adj}}=.002}$) and Behavioral Intention to Use ($t(9)=5.238$, $\boldsymbol{p_{\text{adj}}=.003}$) (see~\cref{fig:result-sus-tam}-B).
No significant differences were found for Perceived Ease of Use ($t(9)=1.832$, $p_{\text{adj}}=.3$) or Trust ($t(9)=1.222$, $p_{\text{adj}}=.506$), nor for Satisfaction (Wilcoxon signed-rank test; $W=4$, $p_{\text{adj}}=.85$).
This suggests that IGUANA is perceived as more useful with higher intent to adopt, while ease of use, trust, and satisfaction are comparable to RC.

For Task~A, we first analyzed completion time and the average closest distance to waypoints as separate metrics.
Paired t-tests showed no significant differences between the 3D map interface and RC for either completion time ($t(9)=0.195$, $p_{\text{adj}}=1$) or mean distance ($t(9)=2.097$, $p_{\text{adj}}=.195$).
For the photo-taking subtask, IGUANA achieved an 80\% success rate, while only three of ten participants succeeded with RC.
We then examined the relationship between these two metrics using Pearson correlation.
For IGUANA, a strong, significant negative correlation ($r=-0.693$, $\boldsymbol{p=.026}$) showed that a longer, more deliberate strategy was linked to greater accuracy (shorter mean distance).
This downward trend is visible in the scatter plot (see~\cref{fig:result-seq-time-distance}-E).
In contrast, the RC showed no such relationship ($r=-0.278$, $p=.437$) (see~\cref{fig:result-seq-time-distance}-D).
We then analyzed performance scores which we calculated using~\Cref{eq:score_formula}.
A paired t-test did not detect a significant difference in performance scores between IGUANA's 3D map interface and RC ($t(9)=-2.622$, $p_{\text{adj}}=.112$).
The results are shown in a dumbbell plot (see~\cref{fig:result-seq-time-distance}-C), which connects each participant's scores across the two interfaces.

Paired t-tests on the \ac{SEQ} scores revealed opposite results for the two tasks (see~\cref{fig:result-seq-time-distance}-A and B).
Participants found the 3D map interface significantly easier for Task A compared to RC ($t(9)=4.636$, $\boldsymbol{p_{\text{adj}}=.008}$).
In contrast, there was no significant difference in difficulty for the virtual ball interface in Task~B ($t(9)=-2.667$, $p_{\text{adj}}=.156$).

\begin{figure}[t]
  \centering
  \includegraphics[width=\columnwidth]{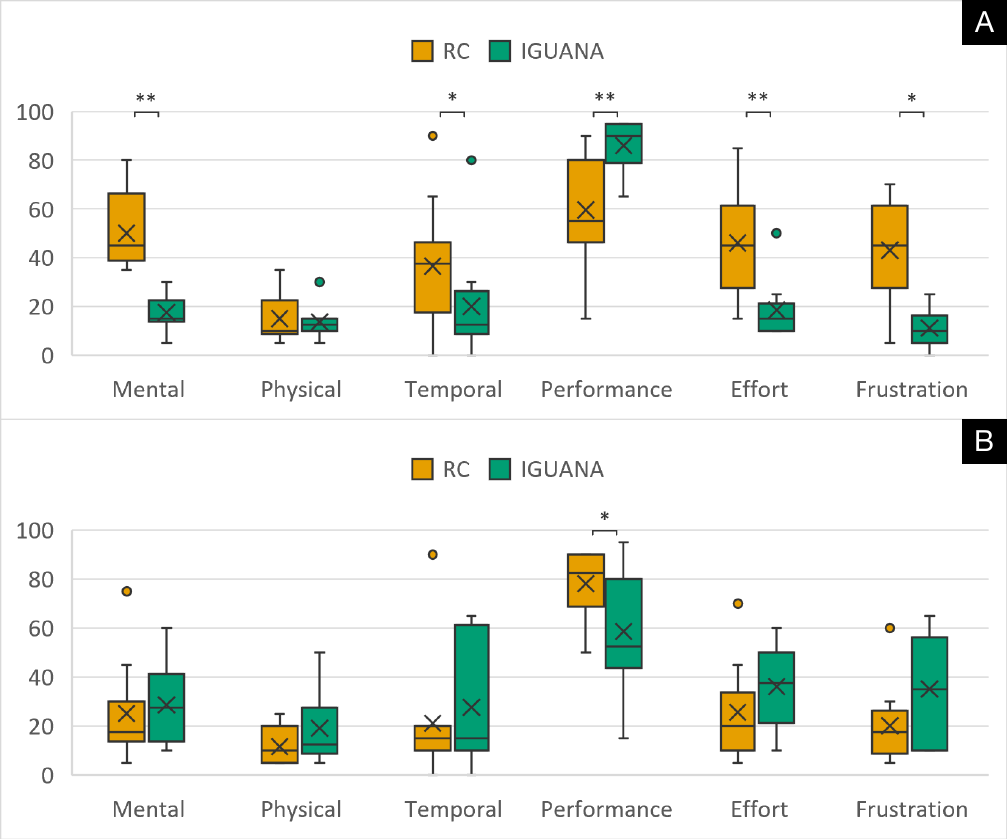}
  \footnotesize Raw Performance scores inverted to maintain consistency in visualization. Significance: * $p_{\text{adj}}<0.05$, ** $p_{\text{adj}}<0.01$.
  \caption{Boxplots of NASA-TLX scores for (A) Task~A and (B) Task~B.}
  \Description{Boxplots of NASA-TLX for Task~A and Task~B under RC and IGUANA conditions.}
  \label{fig:result-tlx}
\end{figure}

For Task~A, paired t-tests on the \ac{NASA-TLX} scores showed that participants experienced significantly lower workload with the 3D map interface compared to RC (see~\cref{fig:result-tlx}-A).
This was observed on Mental Demand ($t(9)=-5.027$, $\boldsymbol{p_{\text{adj}}=.007}$), Temporal Demand ($t(9)=-4.038$, $\boldsymbol{p_{\text{adj}}=.015}$), Effort ($t(9)=-4.793$, $\boldsymbol{p_{\text{adj}}=.008}$), and Frustration ($t(9)=-4.355$, $\boldsymbol{p_{\text{adj}}=.012}$).
They also reported significantly higher subjective Performance with the 3D map interface ($t(9)=4.836$, $\boldsymbol{p_{\text{adj}}=.008}$).
No significant difference was found for Physical Demand (Wilcoxon signed-rank test; $W=14$, $p_{\text{adj}}=1$).

For Task~B, paired t-tests on the \ac{NASA-TLX} scores revealed that the virtual ball interface resulted in significantly lower subjective Performance compared to RC ($t(9)=-3.482$, $\boldsymbol{p_{\text{adj}}=.049}$) (see~\cref{fig:result-tlx}-B).
While other workload ratings were numerically higher for the virtual ball interface, no statistically significant differences were found for Mental Demand ($t(9)=0.506$, $p_{\text{adj}}=1$), Physical Demand ($t(9)=1.8$, $p_{\text{adj}}=.525$), Effort ($t(9)=1.542$, $p_{\text{adj}}=.612$), Frustration ($t(9)=1.562$, $p_{\text{adj}}=.612$), or Temporal Demand (Wilcoxon signed-rank test; $W=10$, $p_{\text{adj}}=1$).

\subsection{Qualitative Results}

A descriptive quality analysis approach was used to organize and summarize the interview results.
By systematically reviewing the feedback, we were able to identify key themes such as the usefulness of the 3D map interface, the usability challenges of the virtual ball interface, and the mixed responses to the spatial overlay.

Participants consistently praised the 3D map interface, describing it as "the best option," "intuitive," and "easy to use."
\textit{P4} remarked, "I like the declarative approach in the 3D map interface better than the RC's imperative style."
Similarly, \textit{P3} liked that "there is no micromanaging with the 3D map interface."
Features such as the height indicator and preview for the camera waypoint were described as helpful.
Overall, the most positively received aspect was the fact that the participants did not need to manually control the \ac{UAV}.

As for waypoint placement, several suggestions were recorded.
\textit{P9} observed that having "a ground shadow for the waypoint marker would make it better" and, along with \textit{P10}, suggested adding movement scaling for finer control.
A separate input for height was also suggested by several participants \textit{(P6, P7, P9)}.

Although all participants understood the intent behind combining multiple axes controls into the virtual ball interface and acknowledged its intuitiveness, some found it harder to use than a standard dual-stick controller \textit{(P8, P10)}.
The highest complaints were the lack of physical feedback from the virtual ball, resulting in them frequently looking down to make sure they were performing the correct gesture.
The visual and auditory feedback alone were not enough to make participants feel confident \textit{(P8)}.
Additionally, a short tutorial session was perceived as insufficient to familiarize them with the hand pose and movement, which is very new to them.
\textit{P10} noted, "Maybe it's easier to grab the ball,” and with \textit{P7} added, "It needs an axis locking option."
However, seven participants believed that with enough exposure, the virtual ball interface could be as easy to use as the RC (\textit{P2, P3, P5, P6, P7, P9, P10}).

All participants agreed that the spatial overlay helped them to quickly locate the \ac{UAV} in the sky.
As \textit{P7} pointed out, "I find myself constantly looking at the map when using the RC."
However, participants expressed mixed opinions regarding the amount and placement of additional information on the overlay.
Two participants did not find it necessary to display anything beyond distance and height \textit{(P1, P5)}.
Three participants suggested moving the additional information closer to the ground \textit{(P4, P5, P9)}.
Additionally, four participants complained about the constant size of the overlay, suggesting that it should scale with the distance \textit{(P4, P5, P7, P10)}.

\section{Discussion}
\label{sec:discussion}

\subsection{IGUANA Proposed Interfaces}

Our user study revealed that the 3D map interface was particularly well-received.
Participants consistently praised the interface for relieving them from any manual control when flying, suggesting that automatic navigation is still preferable to manual control for guiding the \ac{UAV} to follow a predefined path.
They rated it as more usable and useful with higher adoption intent, and also experienced significantly less workload overall.
Performance scores were numerically higher, though this difference was not statistically significant after correction, suggesting that a larger sample may be needed to further investigate this effect.
Using Google's Photorealistic 3D Tiles helps users navigate and understand spatial relationships in the 3D \ac{WIM} reconstructed from real ones~\cite{betancourtExocentricControlScheme2023,sedlmajerEffectiveRemoteDrone2019,sinaniDroneTeleoperationInterfaces2025}.
This resulted in a consistency in trajectory, as all participants produced similar paths using the 3D map interface.
Participants found the 3D map interface intuitive and easy to use in its current form but suggested enhancements for finer control, including: (1)~\textbf{marker movement scaling}, which allows for adjustable, fine-grained control of marker placement, (2)~\textbf{marker ground shadow}, which shows the marker's projected position on the ground level, and (3)~\textbf{separate input for height}, which decouples the position and the height of a marker, allowing precise adjustment for each.

The spatial overlay was considered a primary feature for this type of \ac{MR} application, confirming previous works~\cite{atanasyanAugmentedRealityBasedDrone2023,emamiUseImmersiveDigital2025,liuAugmentedRealityInteraction2020,zollmannFlyARAugmentedReality2014}, particularly when users needed to track moving objects.
Showing the planned waypoint path in the spatial overlay helps users see clearly where the \ac{UAV} is going to go~\cite{walkerCommunicatingRobotMotion2018,zollmannFlyARAugmentedReality2014,emamiUseImmersiveDigital2025}.
Observing the participants' behavior when completing tasks using RC, eight participants looked down at the screen most of the time, reducing their situational awareness~\cite{atanasyanAugmentedRealityBasedDrone2023}.
In contrast, with the IGUANA interface, the spatial overlay was positioned between the user and the \ac{UAV}, allowing more direct visual engagement.
One technical limitation encountered was the sensor precision of the \ac{UAV}'s \ac{GPS} location and height, both of which can drift over time.
To compensate, we implemented a relatively large overlay area.
While this mitigated the drift issue, it also resulted in a less precise location indicator.
Consistent with prior findings ~\cite{atanasyanAugmentedRealityBasedDrone2023}, participants' concerns about the constant overlay size suggest that adaptive scaling could provide a more intuitive visual cue, producing a sense of distance at a glance without needing to read the distance or height information first.

The difficulties participants experienced with the virtual ball interface indicate that physical feedback remains crucial in aiding control~\cite{sehadLocomotionBasedUAVControl2023,ibrahimovDronePickObjectPicking2019,yashinAeroVrVirtualRealitybased2019}, particularly for first-time use.
To improve the virtual ball interface, participants proposed: (1) \textbf{axis locking} control for more precise directional control~\cite{allenspachDesignEvaluationMixed2023}, and (2) \textbf{grab-based interaction} allowing users to grab and hold the ball in hand rather than push it, making an impression that the \ac{UAV} is in the user's hand~\cite{konstantoudakisDroneControlAR2022,betancourtExocentricControlScheme2023}.
While some participants strongly agreed that, with sufficient practice, the virtual ball interface could become as effective as a traditional dual-stick controller, further study is needed to validate this.

\subsection{Limitations}

The small sample size of 10 participants may not capture the full range of user experiences and preferences.
To address this, we employed a within-subjects design, which, however, may have introduced learning effects.
As participants became more familiar with the task over time, this could have potentially skewed the results in favor of the second system used.

The use of a single \ac{UAV} model (DJI Mavic Air) may also limit the generalizability of our findings.
Different \ac{UAV}s may have different control schemes, capabilities, and user interfaces, which could influence user interaction and performance.

Furthermore, this study primarily focused on qualitative assessments of user experience, which, while valuable for understanding subjective impressions, may not fully capture objective metrics such as speed or reliability.

\subsection{Future Work}

Optional haptic feedback could help first-time users feel more confident with the virtual ball interface and could be removed once trust is established.

Future experiments could explore different \acp{UAV} equipped with improved sensors and communication protocols, offering better \ac{GPS} accuracy, more precise altitude estimation, and greater communication range.

Additionally, future user studies should consider adopting a between-subjects design.
This may introduce variability in the results, as different participants may have varying levels of experience and familiarity with \ac{UAV} operation.
However, the effect can be mitigated by having a larger and more diverse participant pool.

\section{Conclusion}
\label{sec:conclusion}

In this paper, we present IGUANA, an \ac{MR}-based interface for \ac{UAV} guidance, navigation, and control.
IGUANA offers (1)~high-level control via a 3D map interface, (2)~low-level control via a virtual ball interface, and (3)~enhanced situational awareness through a spatial overlay.

We conducted a user study to evaluate the usability of IGUANA and found that
(1) the 3D map interface was intuitive and easy to use, suggesting better accuracy and greater consistency with lower perceived workload compared to conventional dual-stick controller,
(2) the virtual ball interface was intuitive but required practice due to the lack of physical feedback, and
(3) the spatial overlay was useful for instantly spotting the \ac{UAV} and enhancing situational awareness.
With further refinement, \ac{MR}-based interfaces could become a viable alternative to conventional \ac{UAV} controllers.

\begin{acks}
  This work was funded by
  \begin{itemize}
    \item the European Social Fund (ESF Plus) and the German Federal State of Saxony within the project ProSECO (100687967),
    \item the European Union, co-financed by tax revenues based on the budget adopted by the Saxon State Parliament, as part of an ESF Plus scholarship (100670474) awarded to T. Krisanty,
    \item the German Federal Ministry of Education and Research (BMBF, SCADS22B) and the Saxon State Ministry for Science, Culture and Tourism (SMWK) by funding the competence center for Big Data and AI \textquotedblleft ScaDS.AI Dresden/Leipzig\textquotedblright, and
    \item DFG grant 389792660 as part of TRR 248 - CPEC.
  \end{itemize}
\end{acks}

\bibliographystyle{ACM-Reference-Format}
\bibliography{references}

\end{document}